\documentclass[twocolumn,aps,superscriptaddress,10pt,showpacs,showkeys,prl]{revtex4-2}
\usepackage{ucs}

\usepackage[utf8x]{inputenc}
\usepackage{graphicx}
\usepackage{setspace}
\usepackage{units}
\usepackage{xspace}
\usepackage{color}
\usepackage[normalem]{ulem}
\graphicspath{{Figs/}}

\begin{document}

\title{Noncollinear spin structure in Dy-doped classical ferrimagnet}

\author{Anupam K. Singh}
\affiliation{Max Planck Institute of Microstructure Physics, Weinberg 2, 06120, Halle (Saale), Germany}

\author{Katayoon Mohseni}
\affiliation{Max Planck Institute of Microstructure Physics, Weinberg 2, 06120, Halle (Saale), Germany}         

\author{Verena Ney}
\affiliation{Institute for Semiconductor and Solid State Physics, Johannes Kepler University Linz, Altenberger Straße 69, 4040 Linz, Austria}

\author{Andreas Ney}
\affiliation{Institute for Semiconductor and Solid State Physics, Johannes Kepler University Linz, Altenberger Straße 69, 4040 Linz, Austria}    

\author{Yicheng Guan}
\affiliation{Max Planck Institute of Microstructure Physics, Weinberg 2, 06120, Halle (Saale), Germany}

\author{Ilya Kostanovski}                                      \affiliation{Max Planck Institute of Microstructure Physics, Weinberg 2, 06120, Halle (Saale), Germany} 

\author{Malleshwararao Tangi}                                 \affiliation{Max Planck Institute of Microstructure Physics, Weinberg 2, 06120, Halle (Saale), Germany}

\author{Mostafa I. S. Marzouk}                                  \affiliation{Max Planck Institute of Microstructure Physics, Weinberg 2, 06120, Halle (Saale), Germany}

\author{Manuel Valvidares}
\affiliation{ALBA Synchrotron Light Source, E-08290 Cerdanyola del Valle`s, Barcelona, Spain}

\author{Pierluigi Gargiani}
\affiliation{ALBA Synchrotron Light Source, E-08290 Cerdanyola del Valle`s, Barcelona, Spain} 

\author{Jean-Marc Tonnerre}
\affiliation{Institut Neel, CNRS et Universite Joseph Fourier, BP. 166, 38042 Grenoble Cedex 9, France}

\author{P. F. Perndorfer}
\affiliation{Institute for Theoretical Physics, Johannes Kepler University, Altenberger Strasse 69, A-4040 Linz, Austria}
\affiliation{Department of Engineering and Computer Sciences, Hamburg University of Applied Sciences, Berliner Tor 7, D-20099 Hamburg, Germany}

\author{P. A. Buczek}
\affiliation{Department of Engineering and Computer Sciences, Hamburg University of Applied 
             Sciences, Berliner Tor 7, D-20099 Hamburg, Germany}            

\author{Arthur Ernst}
\affiliation{Institute for Theoretical Physics, Johannes Kepler University Linz, Altenberger Straße 69, 4040 Linz, Austria}
\affiliation{Max Planck Institute of Microstructure Physics, Weinberg 2, 06120, Halle (Saale), Germany}
\affiliation{Donostia International Physics Center (DIPC), 20018 Donostia-San Sebasti\'{a}n, Spain}

\author{Holger L. Meyerheim}
\affiliation{Max Planck Institute of Microstructure Physics, Weinberg 2, 06120, Halle (Saale), Germany}

\author{Stuart S. P. Parkin}
\affiliation{Max Planck Institute of Microstructure Physics, Weinberg 2, 06120, Halle (Saale), Germany}

\date{\today}

\begin{abstract} 
Noncollinear spin structures have attracted tremendous attention because they offer a versatile platform for spin control and manipulation, essential in spintronics. Realizing noncollinearity in ferrimagnetic insulators is of particular interest as they can be potentially utilized in low-damping spintronics with tunable magnetic order. Within the spinel-ferrite family, Zn- and Al-substituted nickel ferrite (NiZAF) has emerged as an excellent choice for low-damping spintronics. However, realizing noncollinearity in such systems remains challenging. Here, we present evidence of noncollinear spin structure in the NiZAF thin films induced by the rare earth Dy-doping, utilizing soft x-ray spectroscopy methods such as magnetic circular dichroism and x-ray resonant magnetic reflectivity (XRMR). In particular, XRMR reveals a spiral-type spin structure which is attributed to the Dzyaloshinskii-Moriya interaction, arising due to broken inversion symmetry by the Dy-induced local strain field as confirmed by our theoretical calculations. The realization of noncollinearity in the spinel-ferrite opens a pathway to explore the possibility of chiral magnetic domains and topological spin textures exhibiting promise for oxide-based spintronics.  
\end{abstract}


\maketitle
Recent decades witnessed vast attention to the noncollinear spin structures that have been claimed to host great potential, especially for spintronics applications~\cite{fert2017magnetic}. Various types of noncollinear spin structures, such as spiral, conical, and emergent topological spin textures have been identified in several metallic and oxide systems~\cite{fert2017magnetic, trier2022oxide, yang2021chiral}. It has been realized that such spin textures exhibit tremendous potential for fast and energy-efficient data processing and manipulation in novel spintronics which highly relies on the spin-currents required to control spin orientation and manipulation enabled through the interplay of charge and spins~\cite{yang2021chiral, velez2019high}. Nevertheless, there are some key limitations with metals for example, it is often difficult to create pure spin-currents in the metals due to concomitant charge current leading to high-damping and considerable power loss~\cite{Emori2017}. In this respect, magnetic oxide insulators have been noticed as a potential alternative that can be employed to generate pure spin-currents in the adjacent layer with low-damping and dissipation~\cite{trier2022oxide}. Moreover, in comparison to metals, a better chemical and structural tunability of oxides allows better control and manipulation of the desired properties~\cite{trier2022oxide, yang2021chiral, Avci2019}. Particularly, ferrimagnetic rare earth (RE) iron garnets have been claimed as an excellent choice for low-damping spintronics including the stabilization of noncollinear spin structures. However, there are some serious drawbacks of RE garnets in the form of problematic crystal compatibility with other spintronic materials, complicated crystal structure and weak magnetoelastic response, which is undesirable in microelectronics~\cite{emori2021ferrimagnetic}. 

Although classical nickel ferrite has been investigated for decades with its possible utilization in numerous technological applications~\cite{sugimoto1999past}, the Zn/Al-substituted nickel ferrite (Ni$_{0.65}$Zn$_{0.35}$Al$_{0.8}$Fe$_{1.2}$O$_{4}$: NiZAF) has recently been realized as another promising candidate for low-damping spintronics with many advantages over RE garnets~\cite{Emori2017, emori2021ferrimagnetic, Lumetzberger2022}. Ideally, it crystallizes in the inversion symmetric inverse spinel structure (Fd$\overline{3}$m) wherein the spins of cations (Ni$^{2+}$, Fe$^{3+}$, Fe$^{2+}$) in the octahedral sites (O$_{h}$) are aligned antiferromagnetically to those (Fe$^{3+}$) in tetrahedral sites (T$_{d}$), while the resulting moment originates mainly from Ni$^{2+}$, enabling the collinear ferrimagnetic structure~\cite{Emori2017}. However, a detailed study of the spin structure and possibility of fascinating noncollinear spin textures in such systems is still lacking~\cite{bharathi2011structural, hoppe2015enhanced, ugendar2019cationic}. The stabilization of exotic noncollinear spin textures, such as skyrmions has been claimed to be associated with the interplay between the Heisenberg and antisymmetric Dzyaloshinskii-Moriya interaction (DMI)~\cite{Dzy1958, Moriya1960}. For the latter to appear, the absence of inversion symmetry is required, which can be achieved by symmetry breaking in the bulk, e.g., by the presence of defects and atomic relaxations~\cite{Chakraborty2022}. Nevertheless, only a limited number of bulk asymmetric compounds have been observed~\cite{Nayak2017, Srivastava2020}. Another approach is to create the interfacial DMI by taking advantage of the (natural) symmetry breaking at the ferromagnet/heavy metal interface, which is mediated by the strong spin orbit coupling (SOC) of the heavy metal, was shown to generate a large charge to spin current conversion owing to the spin Hall effect RE garnets with heavy metals like Pt~\cite{velez2019high, Avci2019, Caretta2020, Wang2020}. 

Following these previous studies, we move one step further by employing an alternative approach combining the RE-induced high SOC with bulk inversion symmetry breaking. This is achieved by doping an inversion symmetric NiZAF thin films with larger RE Dy$^{3+}$ ions (r=0.91 ~\AA) (with strong SOC) compared to Fe$^{3+}$ and Ni$^{2+}$ ions (r=0.65 and 0.69 ~\AA, respectively~\cite{shannon1976revised}) to distort the crystal structure. Here, we present the evidence of spiral-type noncollinear spin structure in Dy-doped NiZAF thin films using magnetization, Kerr microscopy and x-ray spectroscopy methods. Doping by a few atomic percent with Dy$^{3+}$ induces a local strain field in the vicinity of the occupied O$_{h}$-site, which is attributed to inducing DMI and noncollinearity as confirmed by our first principle calculations. This study opens new perspectives for the strain-gradient engineering of the spin texture in magnetic insulators as potential spintronic systems. 

\textbf{Results and Discussions}

\textbf{1) Crystal structure and magnetic properties}

Details about the sample growth and characterizations are given in the Supplementary Material (SM)~\cite{SUPM}. A typical unit cell of nickel ferrite with inverse spinel structure (Fd$\overline{3}$m) is shown in Fig. 1a, where Fe$^{3+}$ spins occupying the O$_{h}$ nearly compensate for those occupying the T$_{d}$-site, and resultant moments arises from the Ni$^{2+}$ (with a small fraction of  Fe$^{2+}$) occupying the O$_{h}$-site. In the first step, we briefly discuss the key results obtained from our home laboratory. To get depth about the quality of our ultrathin films, XRD is collected using an advanced high-resolution six-circle x-ray diffractometer equipped with a Ga-jet source (more details are given in the SM~\cite{SUPM}). Fig. 1b shows the XRD pattern collected along the perpendicular q$_{z}$ direction in the vicinity of the (311) reflection for the pristine and 5\% Dy-doped NiZAF thin films. In both cases, the high quality of the films are confirmed by the presence of Laue oscillations (marked in Fig. 1b) on both sides of the (broad) film reflection centered near q$_{z}$=0.96 reciprocal lattice units (rlu). The intense and narrow reflection from the substrate is located at q$_{z}$= 1 rlu, indicating that the film has a 4\% expanded c-lattice constant and is tetragonally distorted. The XRD analysis reveals a tetragonally distorted structure in the space group I4$_{1}$/amd (Nr. 141) with lattice parameters of a=b=8.08~\text{\AA}, c=8.32~\text{\AA} for the pristine and a=b=8.08~\text{\AA}, c=8.41~\text{\AA} for the Dy-doped NiZAF.

\begin{figure}[t]
\center
\includegraphics[width=1\linewidth]{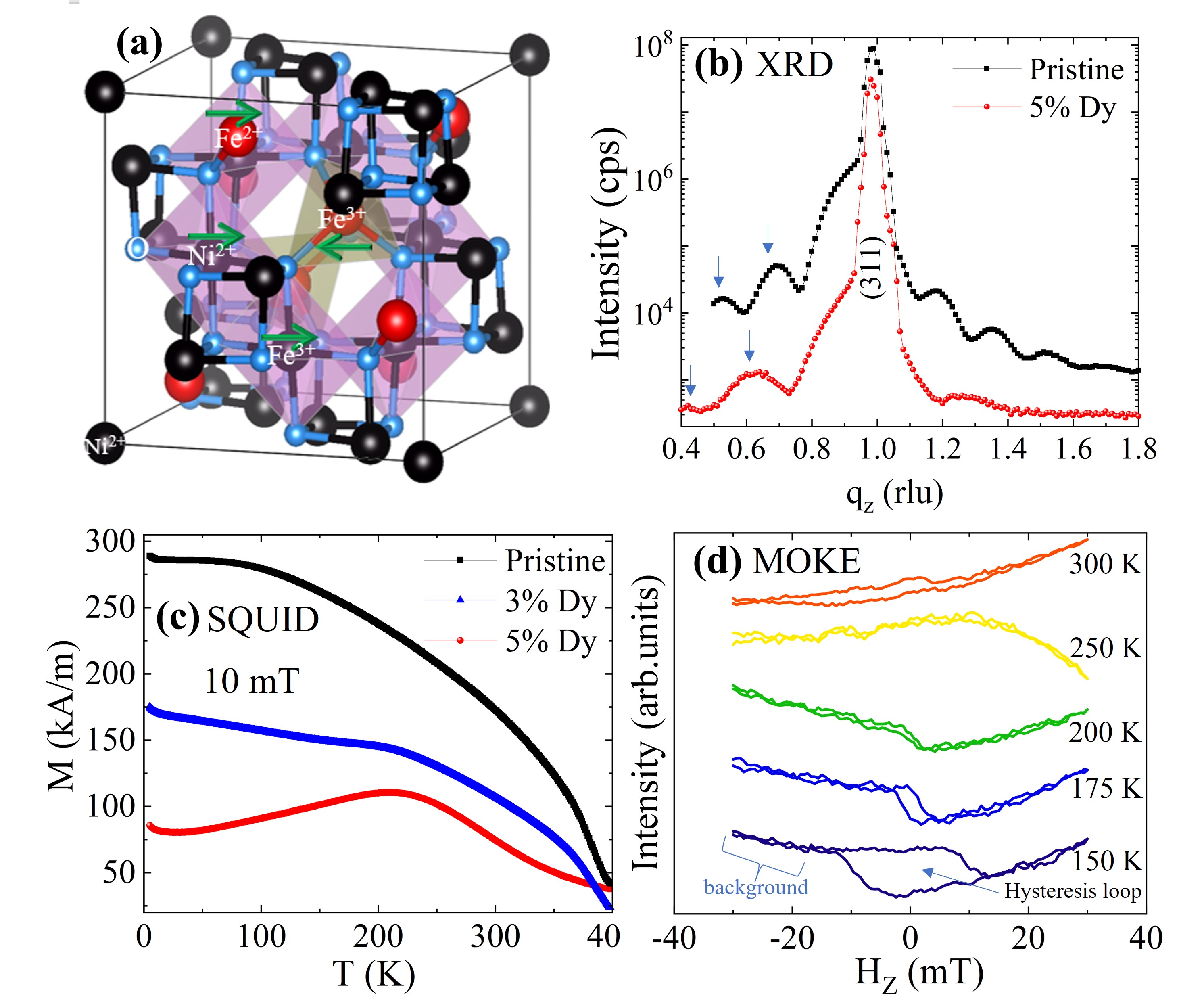}
\caption{\textbf{Crystal structure and magnetic properties.} \textbf{(a)} Unit cell of inverse spinel (Fd$\overline{3}$m) nickel ferrite, where green-colored arrows indicate spin directions at different crystallographic sites such as O$_{h}$ (purple regions) and T$_{d}$ (light green regions). \textbf{(b)} Room temperature XRD pattern collected along the q$_{z}$ direction in the vicinity of the (311) reflection for the pristine and 5\% Dy-doped NiZAF thin films (curves are shifted vertically for clarity), where vertical blue arrows represent Lau oscillations.  
\textbf{(c)} Temperature-dependent magnetization measured using SQUID under the in-plane field of 10 mT during the field-cooling cycle for the pristine, 3\% and 5\% Dy-doped NiZAF thin films. \textbf{(d)} Temperature-dependent Kerr signal (MOKE) versus out-of-plane field (H$_z$) for 5\% Dy-doped NiZAF, where the appearance of hysteresis loop below 200 K is indicated by the arrow.}
\label{Fig1}
\end{figure}

\begin{figure*}[t]
\center
\includegraphics[width=1\linewidth]{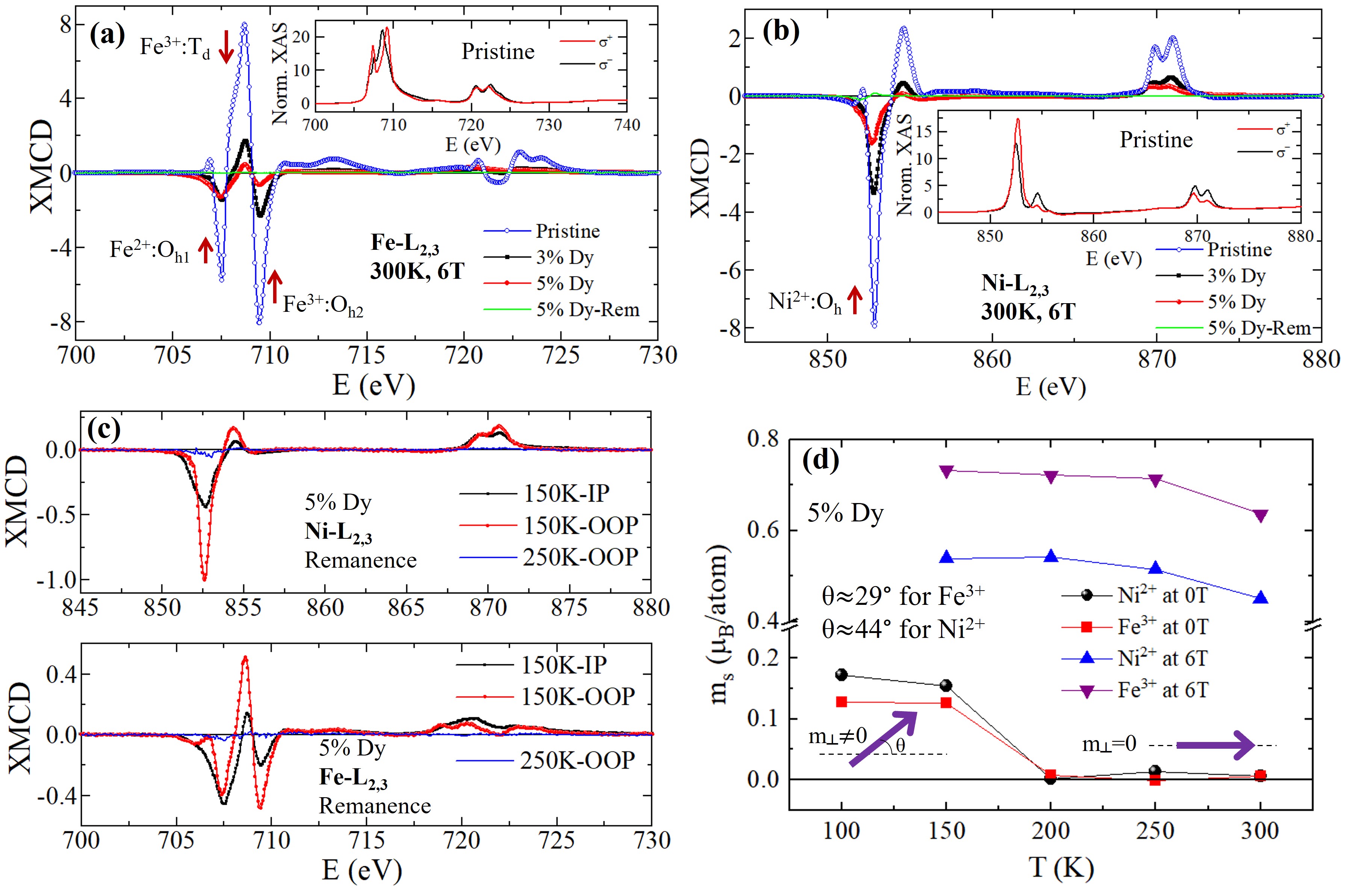}
\caption{\textbf{Noncollinear spin structure using XMCD.} \textbf{(a)} XMCD spectra recorded at 300 K under 6 T using thin films with different Dy concentrations (pristine, 3\% Dy-doped and 5\% Dy-doped) under normal incidence at edges \textbf{(a)} Fe-L$_{2,3}$ and \textbf{(b)} Ni-L$_{2,3}$. Both octahedral (O$_{h1}$ and O$_{h2}$) and tetrahedral (T$_{d}$) sites are labelled, where spins with oxidation state of Fe$^{2+, 3+}$ cations are depicted by red arrows in (a). \textbf{(c)} Comparison of XMCD in the remanence state achieved from the IP and OOP fields for both Ni-L$_{2,3}$ and Fe-L$_{2,3}$ edges for the 5\% Dy-doped NiZAF. \textbf{(d)} Temperature-dependent behavior of the spin magnetic moment (m$_{s}$) for the Ni$^{2+}$ and Fe$^{3+}$ cations, derived from the sum-rule using XMCD spectra collected at corresponding edges in remanence (0 T) and 6 T under normal incidence for the 5\% Dy-doped NiZAF. Purple arrows show a cartoon view of spin-canting below 200 K.}
\label{Fig2}
\end{figure*}

The temperature-dependent magnetization curve measured under an in-plane (IP) magnetic field of 10 mT using SQUID shows two important features as evident by Fig. 1c. The first one is that it shows non-monotonic behavior with a maximum around 200 K for the 5\% Dy-doped NiZAF. This hints as a first indication of modification in the long-range magnetic ordering, in the form of the possible appearance of some out-of-plane (OOP) magnetization component, thereby forming a complex (helical or spin-canted?) spin structure below 200 K. Such features in the magnetization curve have been observed in other noncollinear skyrmion-hosting systems ~\cite{xu2019simultaneous, xiao2019low}. Therefore, we mainly focus on the 5\% Dy-doped NiZAF thin film in this study. The second feature is the significant reduction in the average magnetization value with the Dy-doping as evident by temperature (Fig. 1c) as well as field-dependent hysteresis loops given in Figs. S3a--S3c of the SM. This reflects an enhancement in the antiferromagnetic (AFM) components induced by Dy cations. To get information about possible OOP magnetism in this sample, the Kerr signal that arises through the magneto-optic Kerr effect (MOKE) is recorded with the magnetic field applied perpendicular to the sample plane and the result is shown in Fig. 1d. The appearance of a hysteresis loop in the Kerr signal suggests the emergence of OOP magnetic moments at T $\leq$ 200 K. This manifests the noncollinear or spin-canted magnetic structure at T $\leq$ 200 K in conjunction with temperature-dependent magnetization data for the 5\% Dy-doped NiZAF. The OOP magnetic component below 200 K is also evident from the magnetic hysteresis loop measured using SQUID at 150 K, which is given in Fig. S3e of the SM. A detailed study about this OOP magnetism and noncollinearity is provided in the following sections using XMCD, XRMR and EXAFS data analysis.

\textbf{(2)	X-ray Magnetic Circular Dichroism (XMCD): Magnetism Analysis}

To get a detailed insight about the aforementioned noncollinearity indication appearing in the magnetization and Kerr signal, XMCD data is collected at the BOREAS beamline of ALBA~\cite{BOREAS2016}. We begin the discussion with XMCD recorded at room temperature under a magnetic field of 6 T applied perpendicular to the sample plane in the normal beam incidence and shown in Fig. 2a and Fig. 2b for the Fe-L$_{2,3}$ and Ni-L$_{2,3}$ edges, respectively. Such XMCD signals are obtained from the normalized x-ray absorption spectra (XAS) given in the inset of Figs. 2a,2b. More details about the measurements and analysis procedures are given in the SM. It is evident from Figs. 2a,2b that the intensity of XMCD signal drops significantly for the Dy-doped samples compared to the pristine sample. We note that the XMCD intensity can be quantified by calculating the area under the XMCD peak/dip, which shows a clear drop with Dy-doping as can be seen in Fig. S4a of the SM. This suggests that Dy cations enhance the AFM coupling in the NiZAF, resulting in a drop of average magnetization magnitude as observed from the SQUID measurements (see Fig. 1a and Figs. S3a--S3c of the SM). Moreover, XMCD at the Dy-M$_{4,5}$-edge at 300 K under a 6 T field is shown in Fig. S4f, which shows a typical behavior for Dy located in oxygen environment as previously reported in the Dy-doped nickel ferrite bulk~\cite{ugendar2019cationic}. Note that we did not observe any considerable change in the intensity of the Dy-XMCD signal with increasing Dy concentration (see Fig. S4f of the SM).

\begin{figure*}[t]
\center
\includegraphics[width=1\linewidth]{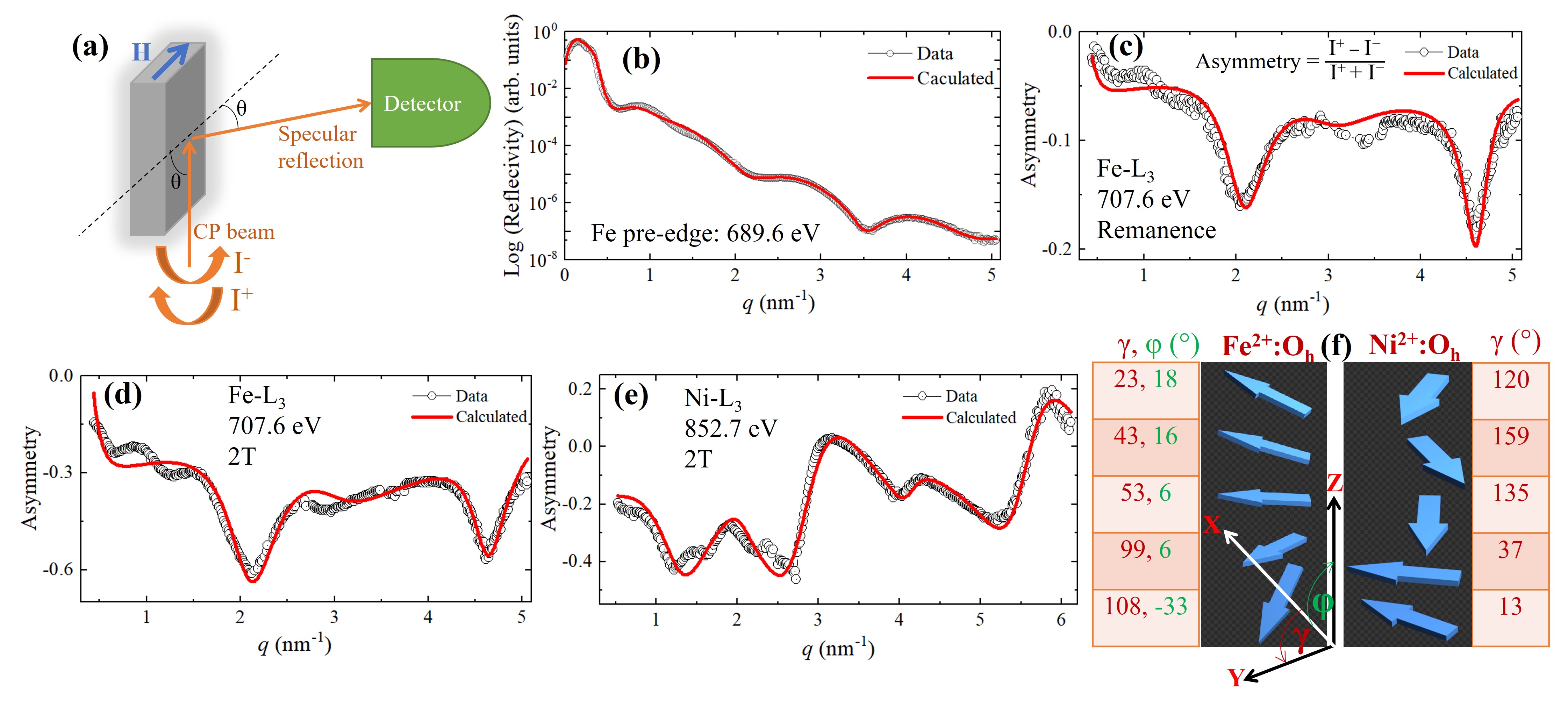}
\caption{\textbf{Spiral-type spin structure using XRMR.} \textbf{(a)} Schematic of the XRMR experimental setup, where specular reflection is detected after the the left(I$^-$)/right(I$^+$) circularly polarized (CP) x-ray beam incidence under the longitudinal geometry with field direction along the sample's plane (indicated by blue arrow). \textbf{(b)}. Magnetic asymmetry modelling at 50 K for the Fe-L$_{3}$ edge \textbf{(c)} in the remanence state and \textbf{(d)} 2 T. \textbf{(e)} Asymmetry modelling at 50 K for the Ni-L$_{3}$ edge at 2 T. \textbf{(f)} Schematic of spin structure for Fe$^{2+}$ (obtained from asymmetry modelling in remanence state) and Ni$^{2+}$ (obtained from asymmetry modelling at 2 T) occupied at the O$_{h}$-site, where the obtained value of IP angle $\gamma$ and OOP angle $\phi$ are given (represented in the coordinate axes model).} 
\label{Fig3}
\end{figure*}

Now we move to the XMCD measured in the remanence state, which is achieved by setting the magnetic field to zero after applying 6 T in the OOP/IP, and subsequently XAS is recorded under the normal/grazing incidence of the beam, thereby the measured XMCD signal is sensitive to the OOP/IP magnetic components in the sample. We observed that the Ni and Fe-XMCD signal measured in the remanence state varies significantly with change in temperature for the 5\% Dy-doped NiZAF. For instance, the XMCD signal in the OOP is found to be vanished above 200 K as can be seen by the XMCD signal at 300 and 250 K given in Figs. 2a-2c. In contrast, significant XMCD in the OOP is detected below 200 K such as 150 K, as can be seen in Fig. 2c. Since we did not observe a measurable OOP XMCD signal in the remanence state at T $\geq$ 200 K, a comparison of IP and OOP XMCD at 150 K is shown in Fig. 2c for Ni and Fe-L$_{2,3}$ edges. Although the presence of an IP XMCD signal is usually expected for a typical spinel-ferrite hosting an IP magnetic anisotropy~\cite{Emori2017, Lumetzberger2020}, the presence of the OOP XMCD signal in the remanence state confirm Dy-doped NiZAF possesses both IP and OOP magnetic components at 150 K (see Fig. 2c). Triggered by this, we performed a detailed sum-rule~\cite{van2014x} analysis to check the behavior of the OOP magnetic moment with change in temperature. A qualitative comparison of temperature-dependent OOP spin magnetic moment (m$_{s}$) derived from the XMCD signal measured in remanence (0 T) and saturation (6 T) is given in Fig. 2d. The standard deviation in the value of m$_{s}$ is estimated to be within 10\%. It is evident from Fig. 2d that OOP m$_{s}$ remains nearly zero above 200 K, due to the IP magnetic anisotropy as expected in such typical spinel-ferrite~\cite{Emori2017, Lumetzberger2020}. However, it increases sharply below 200 K, manifesting the spin-canting from the IP easy axis to the OOP below 200 K for both Ni and Fe spins. Since the magnitude of IP and OOP m$_{s}$ are known from the XMCD signal, the spin-canting angle ($\theta$) can be estimated using the relation tan$\theta$=m$_{\perp}$/m$_{\parallel}$, where m$_{\perp}$ and m$_{\parallel}$ correspond to the OOP m$_{s}$ and IP m$_{s}$ values, respectively. In the remanence state at 150 K, the XMCD provides m$_{\perp}$ ≈ 0.126 and m$_{\parallel}$ ≈ 0.226 $\mu$$_{B}$/atom for Fe$^{3+}$ (while for Ni$^{2+}$, m$_{\perp}$ ≈ 0.154 and m$_{\parallel}$ ≈ 0.162 $\mu$$_{B}$/atom). Using these values, the $\theta$ is estimated to be about 29° for Fe$^{3+}$ and 44° for Ni$^{2+}$. This indicates an unequal canting angle for Fe and Ni spins, and thereby forming a noncollinear spin structure in the 5\% Dy-doped NiZAF. 
At a particular temperature of 150 K in the noncollinear phase, the magnitude of remanence m$_{s}$ is estimated to be about 17\% of saturated m$_{s}$ for Fe$^{3+}$ and 28\% of saturated m$_{s}$ for Ni$^{2+}$. The present value of saturated m$_{s}$ for Ni$^{2+}$ appears in good agreement with the RE-doped nickel ferrite bulk sample reported previously~\cite{ugendar2019cationic}. However, we note the higher value of saturated m$_{s}$ for Fe$^{3+}$ compared to literature~\cite{ugendar2019cationic} is due to the fact that the contribution of the Fe$^{2+}$ at the O$_{h}$-site dominates in our 5\% Dy-doped sample, as discussed above in the context of Figs. S4a-S4d of the SM. It is important to emphasize here that IP m$_{s}$ (i.e., m$_{\parallel}$) are larger than OOP m$_{s}$ (i.e., m$_{\perp}$) in the remanence state, suggesting the higher component of moment lies in the plane even in the noncollinear phase at 150 K. Nevertheless, the present temperature-dependent behavior of OOP m$_{s}$ in the remanence lies in good agreement with our temperature-dependent magnetization and Kerr data (shown in Fig. 1). The emergence of such noncollinearity (or spin-canting) below about 200 K in Dy-doped NiZAF is analogous to those obtained in other noncollinear skyrmion-host systems~\cite{xu2019simultaneous, xiao2019low}. In contrast to Ni and Fe-XMCD, a negligible XMCD signal is observed for the Dy-M$_{4,5}$-edge in the remanence state and remains independent of the temperature change, as can be seen in Figs. S4f,g of the SM, indicating the Dy is not coupled magnetically in the NiZAF lattice. This is also evidenced by the field-dependent XMCD(H) hysteresis loop measured at 100 K for the Dy-O$_{h}$ sites as shown in Fig. S3h of the SM, which shows a typical linear paramagnetic behavior and did not saturate even up to a field range of ±6T. On the other hand, the XMCD(H) hysteresis loops measured for Fe and Ni rapidly saturate at the low fields under both the IP and OOP fields, as evident in Figs. S3f and S3g of the SM. This confirms the presence of both IP and OOP magnetic components with a delicate (or nearly absent) magnetic anisotropy at 100 K in the Dy-doped NiZAF thin film. 

\begin{table*}[htbp]
\centering
\renewcommand{\arraystretch}{1.2}
\begin{tabular}{|c|c|c|c||c|c|c|c|}
\hline
\multicolumn{4}{|c||}{Fe-L$_3$, 50 K, Remanence} &
\multicolumn{4}{c|}{Ni-L$_3$, 50 K, 2T} \\
\hline
No. of layer & Thickness (\AA) & OOP $\phi$ (deg) & IP $\gamma$ (deg) &
No. of layer & Thickness (\AA) & OOP $\phi$ (deg) & IP $\gamma$ (deg) \\
\hline
1/top & 4.5 & 18  & 23  & 1/top & 7.0 & 0 & 120 \\
\hline
2     & 6.0 & 16  & 43  & 2     & 9.5 & 0 & 159 \\
\hline
3     & 5.9 & 6   & 53  & 3     & 6.1 & 0 & 135 \\
\hline
4     & 4.8 & 6   & 99  & 4     & 6.2 & 0 & 37  \\
\hline
5/interface & 4.8 & -33 & 108 &
5/interface & 4.8 & 0 & 13 \\
\hline
\end{tabular}
\caption{Structural parameters obtained from the magnetic asymmetry refinements at 50 K for Fe-L$_3$ and Ni-L$_3$ edges.}
\end{table*}

\textbf{(3)	X-ray Resonant Magnetic Reflectivity (XRMR): Magnetic Asymmetry Analysis}

Motivated by the interesting features of noncollinearity that appeared in the XMCD (Fig. 2) and magnetization data (Fig. 1), we have carried out XRMR experiments~\cite{tonnerre2012depth} to obtain the depth-resolved spin structure of the 5\% Dy-doped NiZAF thin film. The reflectivity data is recorded by detecting the specular reflection in the longitudinal geometry mode at a fixed applied magnetic field (or without field) with switching the beam polarization as shown in Fig. 3a. Raw XRMR data along with deduced magnetic asymmetry (R) at different temperatures (50 and 300 K) and magnetic fields (2 T and remanence) collected at the Fe, Ni-L$_{3}$ and Dy-M$_{5}$ edges are given in Fig. S5 of the SM (see the related discussion in the SM for more details). We note that no considerable feature was observed in the asymmetry at the Dy-M$_{5}$ edge (see Fig. S5f of the SM). Therefore, we focus here on the Fe, Ni-L$_{3}$ edges. In the first step, the reflectivity is modelled using data measured at the energy far before the resonance, as shown in Fig. 3b, and corresponding details are described in context of Fig. S6 of the SM. Subsequently, average reflectivity [I$_{avg}$ = (I$^{+}$+I$^{-}$)/2, where I$^{+}$ and I$^{-}$ correspond to the right and left circularly polarized x-ray beams) is modelled at the resonance energy for Fe-L$_{3}$ and Ni-L$_{3}$ edges, which provide almost similar results to those obtained from the off-resonance reflectivity analysis. After obtaining the geometrical structure through the reflectivity analysis, magnetic asymmetry R is modelled, as shown in Figs. 3c-3e, where IP azimuthal angle ($\gamma$) and OOP polar angle ($\phi$) angles are varied, while geometrical parameters were kept fixed to those values obtained from the reflectivity refinements. Note that in Fig. 3, R at 50 K is shown in the remanence and at 2 T for Fe-L$_{3}$ (Fig. 3c and Fig. 3d), while it is shown only under 2 T for Ni-L$_{3}$ (Fig. 3e) due to fact that no considerable R is observed in the remanence state for the Ni-L$_{3}$ as evident by Fig. S5e of the SM. 

First, we discuss the result obtained from the asymmetry analysis for Fe-L$_3$ edge under remanence. After careful refinement, a total of five magnetic layers of Dy-doped NiZAF were obtained with the thickness of about 2.6(4)~nm, corresponding to the best fit shown in Fig.~3c. The obtained angles $\gamma$ and $\phi$ along with the thickness of individual layers are given in Table~I. For the R modelling, the value of magnetic moment for each Dy-doped NiZAF layer at 50~K is scaled to the average magnetic moment value obtained from the XMCD measured at 150~K through the sum-rule analysis, which is $\approx 0.75~\mu_{\mathrm{B}}$/atom and $\approx 0.14~\mu_{\mathrm{B}}$/atom for Fe$^{3+}$ at 6~T and remanence, respectively. This is enabled by the magnetic moment scaling (mms) parameter in the Dyna~\cite{elzo2012x} software corresponding to the average magnetization scale. The mms were constrained for each layer during the asymmetry modelling and appear to be mms = 1.2, where mms = 1 is equivalent to an average magnetic moment of $\approx 0.14~\mu_{\mathrm{B}}$/atom for Fe$^{3+}$, which is obtained from the XMCD at 150~K in the remanence state. It is evident from Table~I that $\gamma$ changes gradually from the top-most layer ($\gamma = 23^\circ$) to the interface layer ($\gamma = 108^\circ$), indicating a spiral-type noncollinear spin structure in the IP. The $\gamma = 90^\circ$ corresponds to a longitudinal spin configuration along the direction of the external IP magnetic field of 2~T applied before reaching the remanence state. On the other hand, some fraction of magnetization is also observed in the OOP, which is represented by the value of $\phi$ in the asymmetry analysis. A gradual change in the value of $\phi$ from the top-most layer ($\phi = 18^\circ$) to the interface layer ($\phi = -33^\circ$) is obtained, indicating a spiral-type non-coplanar spin configuration in the OOP. The present evolution of the OOP $\phi$ is attributed to the OOP magnetic component observed in SQUID (Figs.~1c, S3e), MOKE (Fig.~1d), and XMCD (Fig.~2d, Figs.~S4f--S4g) measurements. We note that since the present XRMR is collected up to the maximum possible angle of $45^\circ$ (or $q_{\mathrm{max}} = 5~\mathrm{nm}^{-1}$), it may be less sensitive to the OOP component of the magnetization due to only a weak contribution to the $m_z \sin(\theta)$ term in the scattering factor~\cite{tonnerre2012depth}. However, there is a finite value of magnetic asymmetry at $q_{\mathrm{max}}$ at 50~K compared to 300~K (see Fig.~S5c), which can be sufficient to draw a qualitative conclusion in view of the scattering factor~\cite{tonnerre2012depth, meyerheim2009new, violbarbosa2012inhomogeneous}. We note that the present magnetic model is found to be unique, as other models were found to have failed to capture the asymmetry data appropriately, as evident from Fig.~S7 of the SM.
With application of IP external field of 2 T for Fe, overall shape of asymmetry remains nearly same but amplitude of oscillations is considerably enhanced by about 5-times compared to the remanence (see Fig. S5d), indicating the spin-configuration remains same (i.e., spiral) to remanence state but overall increase in the magnitude of net magnetization due to external magnetic field as expected. This is verified by modelling the asymmetry measured under the IP field of 2 T shown in Fig. 3d, where $\gamma$ changes gradually from top-most layer ($\gamma$ = 31°) to interface layer ($\gamma$ = 120°), while average magnetization scales (mms-parameter) come out to be about 4.7-times to that of remanence. Note that the OOP $\phi$ is found to remained zero through all the layers in case of IP external field of 2 T, which align all the spins within the plane of the thin film sample leading to sensitive of XRMR only to the IP component.

Now moving to the Ni-L$_{3}$ edge for which a total of five magnetic layers of Dy-doped NiZAF are also obtained with a total thickness of 3.3(4) nm and corresponding best fit is shown in Fig. 3e. The small difference in the total thickness for Fe and Ni is within the standard deviation (see the SM for more details). It is evident from Table I that $\gamma$ changes gradually from the interface layer ($\gamma$ = 13°) to the upper layer ($\gamma$ = 159°) just below to the top-most layer indicating a spiral-type noncollinear spin structure evolving from the interface to the top layers in the IP as schematically depicted in Fig. 3d. This is in opposite to the Fe for which the spiral-type evolution of the $\gamma$ appears from the top-most to the interface layer (see $\gamma$ value given in Table I and its schematic in Fig. 3f).  Due to an external field of 2 T, $\phi$ remains zero through all layers as the OOP component vanishes at such high IP magnetic field, thereby no OOP magnetization component is observed in the corresponding XRMR measured at 2 T. We now proceed to the comparison of IP $\gamma$ angles for Fe and Ni obtained from magnetic asymmetry analysis. The schematic representation of the noncollinear spin structure for both elements are depicted in Fig. 3f, which shows Ni-spins are not parallel to Fe-spins indicating the enhancement in AFM coupling between them. This is attributed to the significant drop in the IP magnetization (Fig. 1c) and XMCD signal [Fig. 2a,b] for the 5\% Dy-doped NiZAF relative to the pristine sample. We note that pristine nickel ferrite has been reported to exhibit IP collinear Neel-type ferrimagnetic spin-configuration with AFM alignment between T$_{d}$ and O$_{h}$-sites~\cite{ugendar2017effect, yafet1952antiferromagnetic, chappert1967mossbauer}, while triangular-type Yaffet-Kittel spin structure has been realized in Cr-doped nickel ferrite by Mossbauer measurements~\cite{ugendar2017effect, chappert1967mossbauer}. Our XRMR and XMCD results advance one crucial step further and show a noncollinear spin structure with spiral-type evolution of triangular spin-configuration in the IP including some weak OOP component that also shows spiral-type evolution through the depth of Dy-doped NiZAF thin film.

\begin{figure}[t]
\center
\includegraphics[width=1\linewidth]{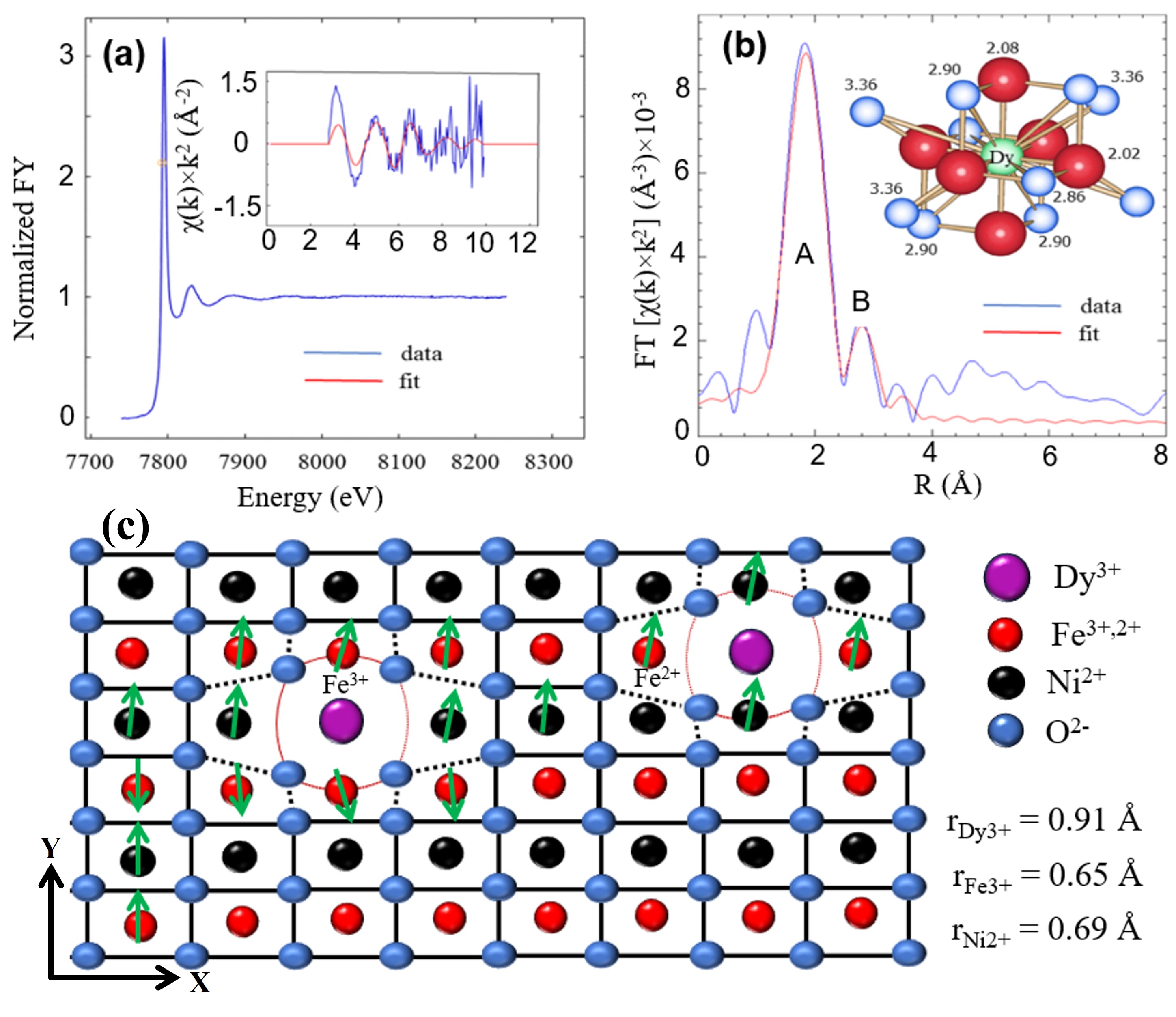}
\caption{\textbf{Local structure and strain-gradient schematic.} \textbf{(a)} Normalized (FY) Dy-L$_3$ edge EXAFS spectrum and $\chi(k)\times k^{2}$ interference function (inset) derived after background subtraction for the 5\% Dy-doped NiZAF. \textbf{(b)} Magnitude of the Fourier Transformation (FT) of $\chi(k)\times k^{2}$. The two maxima labelled by A and B can be related to nearest O and Fe atoms around Dy (see Table II). The inset in (b) shows a model of the local Dy environment within the octahedrally coordinated site. (c) Schematic view of atoms in the lattice showing the strain-gradient induced by larger Dy$^{3+}$ cations, where green arrows show spin directions.}
\label{Fig4}
\end{figure}

\textbf{4)	Extended X-ray Absorption Fine Structure (EXAFS): Local Structure Analysis}

To get an idea about where Dy is sitting in the NiZAF lattice and local structure around Dy, the EXAFS data is collected around Dy-L$_{3}$ edge at the beamline SAMBA of the synchrotron SOLEIL~\cite{belin2005samba}. Fig. 4a shows the EXAFS spectrum after background removal and normalization for the 5\% Dy-doped NiZAF thin film. The detailed analysis was carried out by background subtraction and calculation of the Fourier Transform (FT) of the k$^{2}$-weighted interference function [$\chi$(k)] (see inset) using the WinXAS code~\cite{ressler1998winxas}. The k-range used for the FT-integration extends from $k_{\mathrm{min}} = 2.85~\text{\AA}^{-1}$ to $k_{\mathrm{max}} = 9.8~\text{\AA}^{-1}$. The magnitude of the FT is shown in Fig.~4b.We find two peaks labelled by “A” and “B”, which were fitted in R-space using the program FEFF7 based on theoretical scattering amplitudes and phases~\cite{zabinsky1995multiple}. The local environment around Dy in the O$_h$-site is sketched in the inset of Fig.~4b, which consists of six O atoms at a distance ($d$) between 2.02 and 2.08~\text{\AA}, six Fe atoms at a distance between 2.86 and 2.90~\text{\AA}, and another four Fe atoms at $d = 3.36~\text{\AA}$ if the unrelaxed film structure with the film lattice parameters $a = b = 8.08~\text{\AA}$ and $c = 8.32~\text{\AA}$ is taken as reference. The fit results are listed in Table~II. The FT magnitude in the interval between $R = 1.2$ and $3.2~\text{\AA}$ can be fitted by considering the local structure up to the third shell, albeit with significantly enhanced distances (7-13\%) as compared to the unrelaxed model. We emphasize that the narrow maximum “B” is only correctly fitted by considering the two nearby Fe shells. The finite k-integration range leads to a broader peak if interference effects are absent (see peak “A”). In short, we have found direct evidence that Dy$^{3+}$ ions are incorporated into the O$_{h}$-sites of the spinel structure~\cite{ugendar2019cationic, ugendar2017effect} and involve a considerable local strain-gradient in the vicinity of the large Dy cations.

\begin{table}[h]
    \centering
    \begin{tabular}{|c|c|c|c|c|c|c|}
        \hline
        Shell & R\textsubscript{b} (\AA) & N & R (\AA) & N\textsuperscript{*} & $\sigma^2$ (\AA\textsuperscript{2}) & $\Delta$E\textsubscript{0} (eV) \\
        \hline
        O  & 2.02; 2.08 & 6 & 2.33 (1) & 6.0 (*) & 0.009 & 2.5 \\
        Fe & 2.86; 2.90 & 6 & 3.33 (3) & 4.0 (*) & 0.012 & -1.6 \\
        Fe & 3.36 & 4 & 3.59 (3) & 4.0 (*) & 0.012 & 1.94 \\
        \hline
    \end{tabular}
    \caption{Structural parameters for the the 5\% Dy-doped NiZAF. The meaning of the parameters is as follows: R: refined neighbor distance, R$_{b}$: distance in unrelaxed bulk structure, N$^{*}$: effective coordination number, $\sigma$$^{2}$: mean squared relative displacement amplitude, $\Delta$E$_{0}$: shift of absorption edge, Ru: residual in percent. The amplitude reduction factor (S$_{0}^2$) was kept constant at S$_{0}^2$ = 0.95 in all cases. Parameters labelled by an asterisk (*) are kept fixed. Uncertainties are given in brackets.}
    \label{tab:shell_parameters}
\end{table}

It is essential to understand the origin behind the noncollinear spin structure in the Dy-doped NiZAF thin film. Since the RE Dy$^{3+}$ ion has a larger ionic radius~\cite{shannon1976revised} compared to the transition metals Ni$^{2+}$ and Fe$^{3+}$ ions, its partial substitution may break the inversion symmetry of the inverse spinel lattice~\cite{ugendar2017effect, kamala2008magnetocapacitance, bharathi2011structural}. In other words, large-sized Dy$^{3+}$ can distort its neighborhood and create a local strain-gradient as evidenced by enhanced Dy-O bond distances obtained from the EXAFS analysis (Fig. 4a,b and Table II). As depicted in Fig. 4c, such local strain gradients can break the crystal inversion symmetry locally, which can induce local DMI in an average inversion-symmetric lattice. Such local DMI is expected to be responsible for the stabilization of noncollinear/canted spin structure in the Dy-doped NiZAF. For e.g., strain-gradient-induced DMI has been realized in the other inversion symmetric oxide system [La$_{0.67}$Sr$_{0.33}$MnO$_{3}$ (LSMO)], where DMI-stabilized skyrmions and spiral lattices at zero magnetic field are observed~\cite{zhang2021strain}. A significant impact of local strain fields has been recently proposed on the macroscopic properties~\cite{johnson2024impact, zhang2022hidden}. Moreover, recent theoretical calculations also propose local DMI, induced by broken inversion symmetry locally in an otherwise average inversion symmetric structure, plays a vital factor in the stabilization of topological spin textures~\cite{cui2024anatomy, wang2025skyrmionic, moody2025local}. Therefore, our interpretation of local strain-gradient-induced DMI in Dy-doped NiZAF advances in light of recent theoretical reports and experimental findings of local strain-induced DMI in the other systems~\cite{johnson2024impact, zhang2022hidden, cui2024anatomy, wang2025skyrmionic, moody2025local}.

To elucidate the experimental results, extensive first-principles calculations of electronic and magnetic properties of the Dy-doped NiZAF were performed using a self-consistent Green function method within the multiple scattering theory\cite{Hoffmann2020}. A generalized gradient approximation~(GGA) was utilized to account for electron-electron interaction within the density functional theory~\cite{Perdew1996}. The localized Dy $4f$ electrons and $3d$ states of Ni and Fe were treated within a self-interaction correction method as it is implemented within the multiple-scattering theory~\cite{Perdew1981,Lueders_2005}. The choice of this density functional was motivated by previous first-principles studies of nickel ferrites and rare earth compounds~\cite{Szotek2006,Hughes2007}. 

First, electronic and magnetic properties of the host (NiZAF) were studied within the experimental crystalline structure. The composition of NiZAF (Ni$_{0.65}$Zn$_{0.35}$Al$_{0.8}$Fe$_{1.2}$O$_{4}$) was simulated within a coherent-potential approximation\cite{Soven1967, Gyorffy1972}. The electronic structure of the nickel ferrite and NiZAF  are presented in Figs. S8a,b of the SM. The magnetic structure of the NiZAF is found to be ferrimagnetically ordered. Using the magnetic force theorem within a random phase approximation~\cite{Liechtenstein1987,Tyablikov1995,Hoffmann2020} the critical temperature (T$_{C}$) for the magnetic order was estimated to be 495~K which is close to the experimental value of 450~K \cite{Emori2017}. This fact confirms the proper choice of the DFT functional~\cite{Perdew1981, Lueders_2005}. In the next step, Dy atoms were embedded into the NiZAF lattice using a real-space embedded Green function method~\cite{Zeller1979}, therewith the Green function of the host is Fourier transformed into the real-space representation, taking into account proper boundary conditions. The corresponding electronic structure of Dy-doped NiZAF is shown in Fig. S8c of the SM. The main results of real-space calculations are summarized as following:
\begin{enumerate}
    \item Replacing the O$_{h}$ site-occupied Fe with Dy in the NiZAF cubic supercell (i.e., considering the distortion) leads to a change in neighboring spin direction, i.e., noncollinearity. The tilting is found to be not large and appears only around a few degrees. 
    \item Considering the vacancy at O$_{h}$-occupied Fe in the NiZAF cubic supercell leads to an almost similar effect to case 1, but the strength of noncollinearity appeared smaller (tilting angle) compared to case 1 (Dy occupying the O$_{h}$-site).
    \item The strength of DMI is estimated to be very small, with almost the same D/A magnitude for both cases above. Here D is the DMI and A is spin stiffness. The DMI value is found to be increasing by a distortion of Dy atoms in the NiZAF lattice and facilitates the noncollinear ordering in Dy-doped NiZAF in addition to the exchange interaction between all the moments that surround the Dy atom. As per calculations, the moments are found to be noncollinear only in the nearest and the next nearest coordination shells, i.e., the noncollinearity effect appears very local. 
    \item Exchange interaction between Dy-Ni/Fe is very small, i.e., Dy is not magnetically coupled as evident by XMCD data (Fig. S3h and Fig. S4g of the SM). This is due to the fact that the Dy-$4f$ states are strongly localized, and in the first order they are coupled with the neighboring oxygen atoms. 
    \item The value of T$_{C}$ appears almost the same for both NiZAF and Dy-NiZAF, which is in broad agreement with the experimental SQUID data shown in Fig. 1c.
\end{enumerate}

\section{Conclusion}
In summary, we have presented here compelling evidence of noncollinear spin structure in the Dy-doped spinel ferrite thin films using multiple experimental methods. The noncollinearity feature hinted at by the in-house SQUID and MOKE is confirmed by XMCD, which shows spin-canting below a characteristic temperature of 200 K. A spiral-type spin structure is uncovered by the depth-resolved XRMR. The detailed EXAFS analysis suggests such noncollinearity is attributed to the presence of DMI arising due to broken inversion symmetry locally, leading to a strain gradient around octahedral occupied rare earth Dy ions. Our experimental results are corroborated by theoretical DFT calculations. Our findings are consistent with recent theoretical predictions about local symmetry breaking-induced DMI ~\cite{johnson2024impact, zhang2022hidden, cui2024anatomy, wang2025skyrmionic}. The present evidence of noncollinearity in the insulating spinel-ferrite opens a new avenue for exploring the possibilities of chiral magnetic domains and topological spin textures in centrosymmetric spinel-ferrites, holding great promise for oxide-based spintronic applications.       

\section{Acknowledgments}
The authors thank DFG (Mo 4198/2-1) and FWF (I-5384) for funding. M.V. acknowledges funding by Spanish MINECO grant No. FIS2013-45469. ALBA beamtime access via 2015-IHR-MV and official proposal No. 2022025634 and 2023027446 is acknowledged. We acknowledge SOLEIL for the provision of synchrotron radiation facilities, and we would like to thank Dr. Emiliano Fonda for assistance in using beamline SAMBA for the EXAFS experiment accessed under the proposal number 20230101.

\par

\medskip

\bibliographystyle{apsrev}
\bibliography{./Dy-NIZAF}
\small
\end{document}


\section*{Supplemental Material}

\title{Noncollinear spin structure in Dy-doped classical ferrimagnet}

\author{Anupam K. Singh}
\affiliation{Max Planck Institute of Microstructure Physics, Weinberg 2, 06120, Halle (Saale), Germany}

\author{Katayoon Mohseni}
\affiliation{Max Planck Institute of Microstructure Physics, Weinberg 2, 06120, Halle (Saale), Germany}         

\author{Verena Ney}
\affiliation{Institute for Semiconductor and Solid State Physics, Johannes Kepler University Linz, Altenberger Straße 69, 4040 Linz, Austria}

\author{Andreas Ney}
\affiliation{Institute for Semiconductor and Solid State Physics, Johannes Kepler University Linz, Altenberger Straße 69, 4040 Linz, Austria}    

\author{Yicheng Guan}
\affiliation{Max Planck Institute of Microstructure Physics, Weinberg 2, 06120, Halle (Saale), Germany}

\author{Ilya Kostanovski}                                      \affiliation{Max Planck Institute of Microstructure Physics, Weinberg 2, 06120, Halle (Saale), Germany} 

\author{Malleshwararao Tangi}                                 \affiliation{Max Planck Institute of Microstructure Physics, Weinberg 2, 06120, Halle (Saale), Germany}

\author{Mostafa I. S. Marzouk}                                  \affiliation{Max Planck Institute of Microstructure Physics, Weinberg 2, 06120, Halle (Saale), Germany}

\author{Manuel Valvidares}
\affiliation{ALBA Synchrotron Light Source, E-08290 Cerdanyola del Valle`s, Barcelona, Spain}

\author{Pierluigi Gargiani}
\affiliation{ALBA Synchrotron Light Source, E-08290 Cerdanyola del Valle`s, Barcelona, Spain} 

\author{Jean-Marc Tonnerre}
\affiliation{Institut Neel, CNRS et Universite Joseph Fourier, BP. 166, 38042 Grenoble Cedex 9, France}

\author{P. F. Perndorfer}
\affiliation{Institute for Theoretical Physics, Johannes Kepler University, Altenberger Strasse 69, A-4040 Linz, Austria}
\affiliation{Department of Engineering and Computer Sciences, Hamburg University of Applied Sciences, Berliner Tor 7, D-20099 Hamburg, Germany}

\author{P. A. Buczek}
\affiliation{Department of Engineering and Computer Sciences, Hamburg University of Applied 
             Sciences, Berliner Tor 7, D-20099 Hamburg, Germany}            

\author{Arthur Ernst}
\affiliation{Institute for Theoretical Physics, Johannes Kepler University Linz, Altenberger Straße 69, 4040 Linz, Austria}
\affiliation{Max Planck Institute of Microstructure Physics, Weinberg 2, 06120, Halle (Saale), Germany}
\affiliation{Donostia International Physics Center (DIPC), 20018 Donostia-San Sebasti\'{a}n, Spain}

\author{Holger L. Meyerheim}
\affiliation{Max Planck Institute of Microstructure Physics, Weinberg 2, 06120, Halle (Saale), Germany}

\author{Stuart S. P. Parkin}
\affiliation{Max Planck Institute of Microstructure Physics, Weinberg 2, 06120, Halle (Saale), Germany}

\date{\today}

\pacs{61.05.cp, 73.20.At, 71.15.Mb, 79.60.-i}

\maketitle

\textbf{1.	Experimental Methods}

Single crystalline thin films of Zn/Al-substituted nickel ferrite with a nominal thickness of 5 nm are prepared by reactive magnetron sputtering on a MgAl$_{2}$O$_{4}$ (001) spinel substrate as described and characterized elsewhere~\cite{Lumetzberger2020, Lumetzberger2022}. At optimized conditions, the targets of NiFe$_{2}$O$_{4}$, Al and ZnO are sputtered in a triple magnetron cluster with substrate temperature of 650°C to achieve homogeneous composition~\cite{Lumetzberger2020, Lumetzberger2022}. The Zn$^{2+}$ and Al$^{3+}$ are substituted to achieve nominal composition of Ni$_{0.65}$Zn$_{0.35}$Al$_{0.8}$Fe$_{1.2}$O$_{4}$ (NiZAF) estimated by Rutherford backscattering spectroscopy (RBS) performed using 2 MeV He$^{+}$ beam. The Zn$^{2+}$ cations are substituted in place of Ni$^{2+}$ to reduce the Gilbert damping while Al$^{3+}$ is introduced to minimize the lattice mismatch between NiZAF film and substrate~\cite{Emori2017}. Dy doping was subsequently carried out by dosing the Dy to the pristine films in the molecular beam epitaxy (MBE) chamber under ultrahigh vacuum (~10$^{-9}$ mbar) at 650°C for different time (25-50 seconds) to achieve different concentrations (3-5\%) of Dy. For instance, pristine NiZAF film was covered by approximately 2×10$^{15}$ Dy atoms/cm$^{2}$ followed by annealing at 650°C for several minutes. Subsequent RBS analysis yields a composition of Ni$_{0.66}$Zn$_{0.07}$Al$_{0.80}$Fe$_{1.18}$Dy$_{0.34}$O$_{3.96}$ which corresponds to 5\% Dy-doping and indicates some Zn$^{2+}$ loss while the total amount of Ni$^{2+}$, Fe$^{3+}$ and Al$^{3+}$ are nearly constant and Dy$^{3+}$ has been incorporated into the NiZAF lattice. The modification of the slight-stoichiometry suggests a redistribution of the small Fe$^{3+}$ ions from the octahedral (O$_{h}$) sites, which are occupied by the relatively larger Dy$^{3+}$ ions (rDy$^{3+}$=0.91~\text{\AA} versus rFe$^{3+}$ =0.65 ~\text{\AA})~\cite{shannon1976revised}. Direct evidence for Dy$^{3+}$ ions in O$_{h}$ sites come from extended x-ray absorption fine structure (EXAFS) experiments carried out by collecting the Dy-L$_{\beta}$ fluorescence yield versus photon energy above the Dy-L$_{3}$ absorption edge at the beamline SAMBA of the synchrotron SOLEIL, France~\cite{belin2005samba}. On the other hand, direct evidence of redistribution of the Fe$^{3+}$ cations within tetrahedral (T$_{d}$) and O$_{h}$ sites are observed by x-ray magnetic circular dichroism (XMCD), as described later in the next paragraph.

The high-quality nature of the films is confirmed by six-circle x-ray diffractometer (XRD) equipped with a Ga-jet x-ray source and Bruker D8 XRD machine equipped with a CuK$\alpha$ source. The pristine NiZAF films are found to be tetragonally distorted with lattice parameters a=b=8.08 and c=8.32~\text{\AA}. Magnetic characterization is performed using a superconducting quantum interference device (SQUID) magnetometer (Quantum Design, MPMS). The temperature-dependent Kerr signal that arises via the magneto-optic Kerr effect (MOKE) is collected under the perpendicular magnetic field applied to the sample's plane to get idea about the out-of-plane (OOP) magnetic component. A visible LED light source was used for present MOKE measurement. To obtain detailed magnetic information, x-ray absorption spectroscopy (XAS) spectra were recorded under normal and grazing incidence geometry of x-rays (angle between x-ray beam and sample’s surface: 20 deg for grazing incidence) with ($\mu$$_{0}$H = 6 T) and without magnetic field (remanence) for both left- and right-handed circular polarization of x-rays at the BOREAS (BL29) beamline of ALBA synchrotron source, Spain~\cite{BOREAS2016}. The spectra were collected at the Ni/Fe-L$_{2,3}$ and Dy-M$_{4,5}$ edges in the temperature range between 50 K and 300 K with 50 K interval using Hector chamber. The data were collected in both total electron yield (TEY) mode. Several TEY spectra were collected using left and right-circular polarized x-rays for achieving good statistics and subsequently averaged. From these average TEY spectra, off-set before jump is removed to zero followed by post-edge normalization to 1. Finally, XMCD signal is obtained from taking difference between normalized TEY intensity for left and right-circular polarized x-rays. To determine the depth-resolved spin-structure, soft x-ray resonant magnetic reflectivity (XRMR) experiments were carried out using scattering (MaReS) chmaber at the BOREAS beamline of ALBA. The reflectivity was recorded first at the photon energy at non-resonant condition below the absorption edge and subsequently at the resonance condition at the Fe-L$_{3}$, Ni-L$_{3}$ and Dy-M$_{5}$ edges under longitudinal geometry. Data were recorded in the longitudinal geometry under an in-plane magnetic field ($\mu$$_{0}$H = 2 T) and at remanence by reversing the beam helicity (I$^{+}$ and I$^{-}$). Here, I+ and I- represent the reflectivity recorded using a “plus” and “minus” circularly polarized x-rays beam. The data were collected in the specular $\theta$-2$\theta$ mode at 50 K and 300 K using a Keithley’s silicon detector. Subsequently, magnetic asymmetry R=(I$^{+}$-I$^{-}$)/(I$^{+}$+I$^{-}$) is determined. 

\textbf{2.	Phase Purity and Atomic Composition: XRD and RBS}

The XRD pattern along the (001) direction in $\theta$-2$\theta$ scan for the pristine and Dy-doped NiZAF thin films is given in Fig. S1a. A broad (004) reflection of NiZAF film appear close to the intense (004) reflection of MgAl$_{2}$O$_{4}$ substrate. This suggests the epitaxial film growth along (001). We note that this Bragg reflection of the films is broad and weak compared to the substrate reflection due to fact of small thickness (3-5 nm) of all the films. The chemical composition of each samples is estimated by fitting the RBS spectrum as given in Fig. 1b. The detail of RBS analysis for 5\% Dy-doped NiZAF alongwith the chemical composition obtained for each sample is given at the right-side of Fig. S1b.

For more information on the ultrathin film sample quality, the XRD is collected using an advanced high-resolution six-circle x-ray diffractometer equipped with a Ga-jet source for x-ray generation. This is operated at 70 keV and 100 W electron beam power, emitting high brilliance Ga-K$\alpha$ radiation ($\lambda$ = 1.3414 ~\AA) with a beam size of 100 $\mu$m height and 2 mm width focused by Montel optics~\cite{rimmler2023atomic}. Fig. 1b of the main text shows the XRD pattern collected along the perpendicular q$_{z}$ direction in the vicinity of the (311) reflection of the pristine and 5\% Dy-doped NiZAF films. In both cases a film of high quality is observed as concluded on the presence of Laue-oscillations on both sides of the (broad) film reflection centered near q$_{z}$=0.96 reciprocal lattice units (rlu). The intense and narrow substrate reflection is located at q$_{z}$= 1 rlu indicating that the film has a 4\% expanded c-lattice constant and is tetragonally distorted.

\begin{figure*}[t]
\center
\includegraphics[width=1\linewidth]{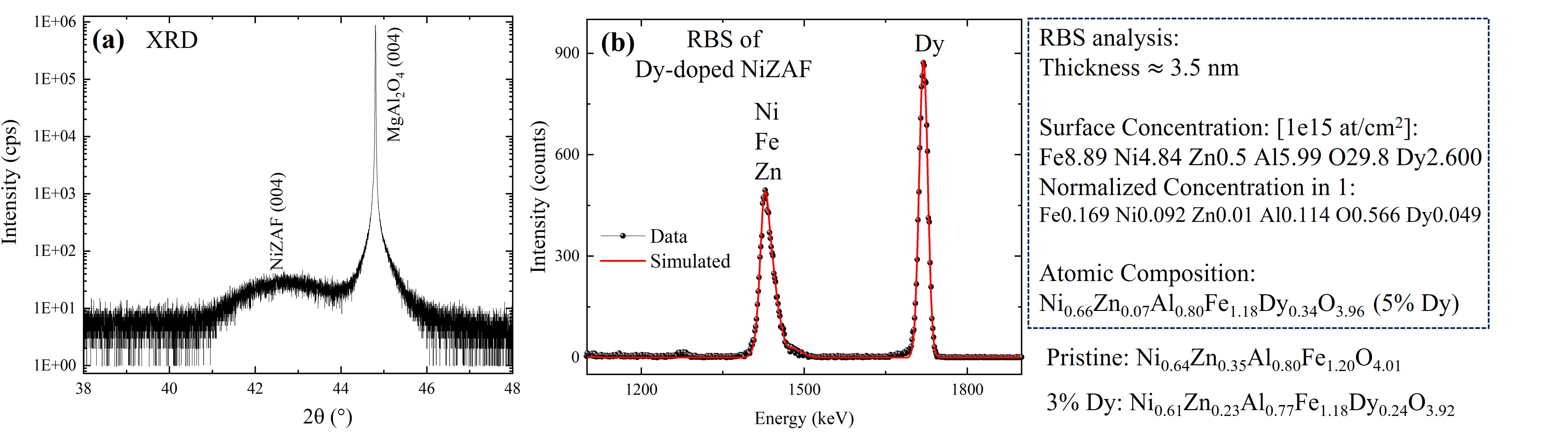}
\caption{\textbf{Structural characterization and atomic composition of the pristine and Dy-doped NiZAF films}. \textbf{(a)} XRD of the pristine NiZAF thin film. (b) Analysis of the RBS spectrum for 5\% Dy-doped NiZAF thin film, where the analyzed thickness and composition are given at the right side of (b). The composition of other samples obtained from RBS analysis is also given at the bottom-right of (b).}
\label{FigS1}
\end{figure*}

To investigate the possibility of structural phase transition in 5\% Dy-doped NiZAF (as indicated by magnetization data described in the next section), low-temperate XRD measurement is performed at 4 K by cooling the sample using liquid helium in the cryostat equipped with a beryllium dome in our six-circle XRD machine. Results of low-temperature XRD are shown in Figs. S2a-S2f. The Lscan in the vicinity of the (111) and (311) Bragg reflections show no extra peak at 4 K compared to 300 K as can be seen in Figs. S2a,b. Further, the Lscan in the vicinity of the (511) Bragg reflection also did not show any extra reflection at 4 K compared to 300 K as can be seen in Fig. S2c, thereby indicating the absence of any structural phase transition. To capture any possible weak reflection, H1Lmesh scan is recorded around the (311) reflection at 4 K, which also did not reveal any additional weak reflections as can be seen in Figs. S2d-S2f.  Thus, the absence of any Bragg peak at 4 K discards any structural phase transition at low temperature in the 5\% Dy-doped NiZAF thin film.

\begin{figure*}[t]
\center
\includegraphics[width=1\linewidth]{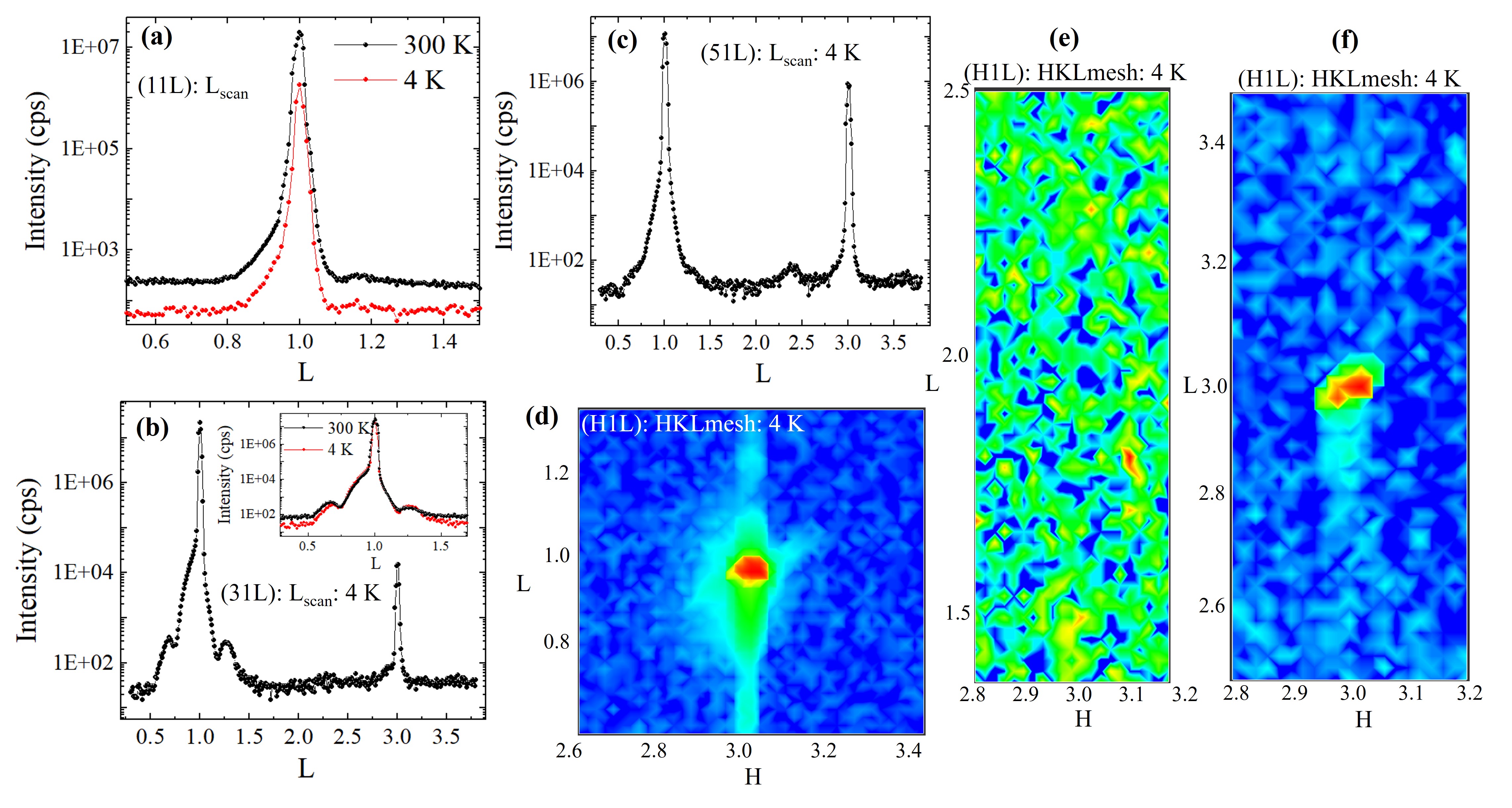}
\caption{\textbf{Possibility of structural phase transition in 5\% Dy-doped NiZAF thin film.} (a) XRD L-scans collected at 300 and 4 K in the vicinity of the (111) Bragg reflection. XRD L-scans collected at 4 K around Bragg reflections (b) (311) and (c) (511). The inset of (b) shows the L-scan around the (311) reflection recorded at 300 and 4 K.  (d)-(f) H1Lmesh XRD scans collected at 4 K around the (31L) reflection. Red color shows maxima of the (311) and (313) Bragg reflections in (d) and (f), respectively.}
\label{FigS2}
\end{figure*}

\textbf{3.	Magnetic Phase Transition: SQUID and MOKE}
Temperature-dependent magnetization [M(T)], measured using SQUID under the in-plane applied magnetic field of 10 mT during the field-cooling (FC) cycle, are shown in Fig. 1c of the main text. The M(T) clearly reveals that the value of magnetization at T = 5 K decreases significantly from around 290 kA/m to 175 kA/m upon 3\% Dy-doping in pristine NiZAF and finally to 86 kA/m for 5\% Dy-doped NiZAF. However, apart from magnetization decrease, the behavior of M(T) for 3\% Dy-doped NiZAF looks almost similar to pristine NiZAF with a very weak hump around 200 K. In contrast, M(T) is deviated significantly with broad peak-feature appeared around 200 K for the 5\% Dy-doped sample. Therefore, 5\% Dy-doped NiZAF is the focus of the present study. At T=5 K, we observe about 70\% reduction of the magnetization from 290 kA/m to 86 kA/m upon 5\% Dy-doping, which can be speculated by Dy$^{3+}$ (M$_{eff}$ = 10.65 $\mu$${B}$) entering the "B" sites (O$_{h}$-sites) and antiferromagnetic coupling to "A" site (T$_{d}$) cations. In contrast to pristine NiZAF for which M(T) follows Bloch’s T$^{-3/2}$ law~\cite{kipferl2003thermal}, the M(T) curve of 5\% Dy-doped NiZAF is non-monotonic with a maximum around T = 200 K (see Fig. 1c in the main text). Since the possibility of structural phase transition is discarded by low-temperature XRD measurements (see Fig. S2), the observed decrease in the in-plane magnetization below 200 K in the FC suggest that some fraction of magnetization is distributed to the OOP i.e., magnetic phase transition. This feature in M(T) is the first indication for the formation of a complex (helical or spin-canted?) spin-structure below 200 K and similar to those obtained in other noncollinear skyrmion-hosting systems~\cite{wang2016centrosymmetric}.

To get more detail about the non-monotonic behavior of M(T) for 5\% Dy-doped sample, Kerr signal were recorded with magnetic field applied perpendicular to the sample plane. This is done to get information about possible OOP magnetism in the sample. It is evident from Fig. 1d of the main text that there is small loop opening in the Kerr signal (arise via MOKE) starting from a certain temperature T = 200 K and this loop intensity gets increases on further reducing temperature for e.g., 175 K and 150 K which is the minimum temperature for data collection in the present study. In contrast, there is no hysteresis loop found above 200 K at temperature 250 K and 300 K, where Kerr signal appears from background only. Note that the signal at higher field (loop closing) are just unphysical background signal. The appearance of hysteresis loop in Kerr signal suggests the emergence of OOP magnetic moments at T $\leq$ 200 K. This manifests the noncollinear magnetic structure at T $\leq$ 200 K in conjunction with M(T) data for the 5\% Dy-doped NiZAF. The detail study about this noncollinearity is provided using XMCD and XRMR data analysis in the upcoming sections and in the main text.

\textbf{4.	Magnetic Hysteresis Loop: SQUID and XMCD}

For pristine and Dy-doped NiZAF thin films, magnetic field-dependent magnetization [hysteresis loop M(H)] measured around low magnetic field (±100 mT) applied under the in-plane (IP) to the sample at 5, 150 and 300 K are shown in Figs. S3a, S3b and S3c, respectively. However, M(H) at 5 K for 5\% Dy-doped NiZAF is measured comparatively higher fields (up to ±400 mT) as M(H) loop closes at relatively at higher field H $>$ ±100 mT. It is evident from these M(H) loop that magnetization value decreases with increasing the Dy concentration. This is in consistent with earlier report on rare earth ion doped nickel ferrite bulk system~\cite{ugendar2019cationic}. Further, although there is no significant change in the coercivity with Dy-doping at 150 and 300 K, it increases considerably at 5 K thereby indicating the hardening the magnetic properties of NiZAF with rare earth element Dy doping. Moreover, a full M(H) loop measured around high field (±5 T) applied to the both IP and out-of-plane (OOP) to 5\% Dy-doped NiZAF film at 300 K and 150 K are shown in Figs. S3d and S3e, respectively. It is evident by Fig. S3d that the sample possess weak IP magnetic anisotropy at 300 K with OOP magnetic anisotropy field of $<$ 0.6 T.  In contrast, the anisotropy between IP and OOP magnetization almost disappears at 150 K (see Fig. S3e) revealing almost equal distribution of magnetic moment in the IP and OOP. This reflects the noncollinear distribution of magnetic moment at 150 K for the 5\% Dy-doped NiZAF sample. 

The XMCD hysteresis loop [XMCD(H)] is also recorded at 100 K under normal as well as grazing incidence of x-rays to the sample plane at the energies Ni-O$_{h}$, Fe-O$_{h1}$ and minima of Dy-XMCD (i.e., Dy-O$_{h}$) for 5\% Dy-doped NiZAF as shown in Figs. S3f-S3h. The grazing incidence and normal incidence to measure XMCD(H) provide information corresponding to the IP and OOP distribution of magnetic components, respectively. It is evident from Figs. S3f, S3g that XMCD signal for Ni and Fe related to the octahedral site saturate quickly at the low field for both IP and OOP in conjunction with the hysteresis loop measured using SQUID (see Fig. S3e) and manifest absence of  magnetic anisotropy between IP and OOP which is in marked contrast with the pristine NiZAF, which hosts in-plane anisotropy~\cite{Emori2017}. This absence of anisotropy hints toward the formation of noncollinear spin structure at 100 K for the 5\% Dy-doped NiZAF sample. Moreover, in contrast to the magnetically coupled Ni and Fe, XMCD(H) for Dy (shown in Fig. S3h) did not saturate even upto ±6 T indicating Dy is not magnetically coupled or it is coupled antiferromagnetically via supercharge interaction through some cations and may require higher field to achieve the saturation state in the XMCD hysteresis loop.

\begin{figure*}[t]
\center
\includegraphics[width=1\linewidth]{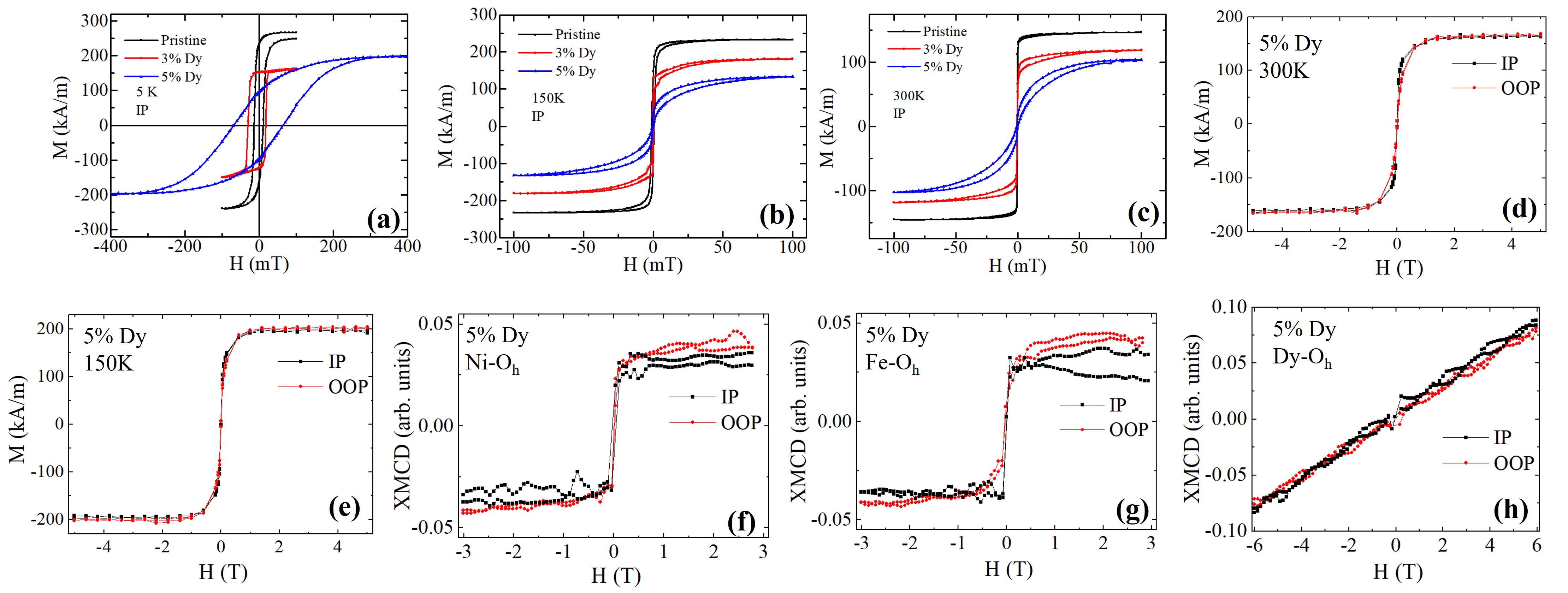}
\caption{\textbf{Magnetic hysteresis loop of pristine and Dy-doped NiZAF films}: M(H) loop measured using SQUID with IP applied magnetic field for pristine and Dy-doped NiZAF at (a) 5 K, (b) 150 K and (c) 300 K. M(H) measured under IP and OOP fields for the 5\% Dy-doped NiZAF at (d) 300 K and (e) 150 K. XMCD hysteresis loop measured at 100 K under normal and grazing beam incidence with magnetic field applied OOP and IP to the sample's plane, respectively, for the 5\% Dy-doped NiZAF at the energies (f) Ni-O$_{h}$, (g) Fe-O$_{h}$ and (h) Dy-O$_{h}$.}
\label{FigS3}
\end{figure*}

\textbf{5.	X-ray Absorption Spectroscopy (XAS): XMCD}

Normalized x-ray absorption spectroscopy (XAS) spectra collected for the pristine and Dy-doped NiZAF thin films using right ($\sigma$$^{+}$) and left ($\sigma$$^{-}$) circular polarized x-rays beam at 300 K under normal beam incidence with 6 T magnetic field applied normal to sample surface at Fe-L$_{2, 3}$ and Ni-L$_{2, 3}$ edges. Such XAS are shown in the inset of Fig. 2a,2b of the main text along with the corresponding XMCD ($\sigma$$^{-}$ - $\sigma$$^{+}$) signal. A well-shaped XMCD signal can be clearly seen. To reveal the effect of Dy concentration on magnetism, XMCD signal for pristine and Dy-doped samples are shown in Figs. 2a,2b of the main text, which shows intensity drop and redistribution in XMCD signal appeared with Dy-doping. Such intensity drop in the XMCD signal is quantified by calculating area under the peak and given in Fig. S4a. Further, such redistribution of Fe-L$_{3}$ XMCD intensity has been realized between magnetite (Fe$_{3}$O$_{4}$) and nickel ferrite (NiFe$_{2}$O$_{4}$)~\cite{richter12009nife2o4}. Although present XAS at both edges and XMCD at the Ni-L$_{2,3}$ edge looks in good agreement with previous report on the pristine NiZAF with different thickness~\cite{Lumetzberger2020}, the behavior of XMCD at the Fe-L$_{3}$ edge is slightly different for the 5\% Dy-doped sample. For a typical spinel-ferrite system, Fe$^{3+}$ is expected to occupy O$_{h2}$-sites with small portions of Fe$^{2+}$ at another O$_{h1}$-site appearing at the lower energy~\cite{Emori2017, kaveev2021high}. Although Fe-L$_{3}$ XMCD shows that the intensity of XMCD signal related to the Fe$^{3+}$ at O$_{h2}$-site is higher than that of the Fe$^{2+}$ at O$_{h1}$-site for the pristine and 3\% Dy-doped sample, a closer look reveals the intensity of the XMCD signal related to the Fe$^{2+}$ at O$_{h1}$-site become higher than Fe$^{3+}$-XMCD at O$_{h2}$-site for 5\% Dy-doped NiZAF as can be seen in Fig. 2a of the main text. This feature is appearing due to the fact that with increasing Dy concentration from 3 to 5\%, the intensity of XMCD signal at O$_{h2}$ and T$_{d}$-sites decreases, while intensity at O$_{h1}$ remains, nearly same as can be seen in Fig. S4a of the SM.  

\begin{figure*}[t]
\center
\includegraphics[width=1\linewidth]{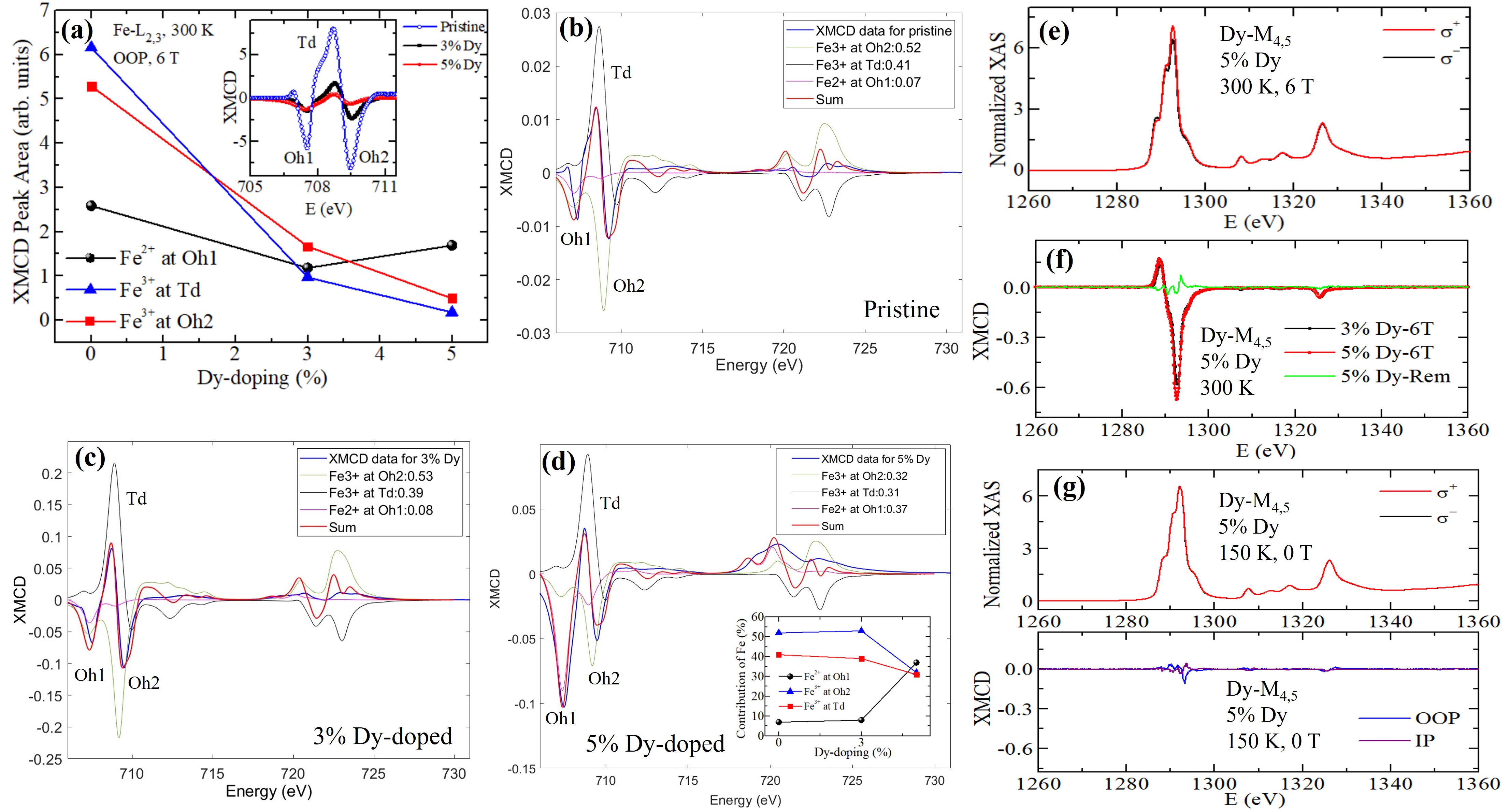}
\caption{\textbf{XMCD analysis for pristine and Dy-doped NiZAF films.} (a) Dy-doping in atomic \% verses area under the XMCD peak (i.e., intensity/dip) at different sites labelled in the inset wherein XMCD for Fe-L$_{3}$ at 300 K, 6 T is shown. Fitting of the XMCD signal for (b) pristine NiZAF (c) 3\% Dy-doped and (d) 5\% Dy-doped. Inset of (d) shows contribution of Fe obtained from the XMCD fitting at different sites. Normalized XAS spectra at the Dy-M$_{4, 5}$-edge for the 5\% Dy-doped NiZAF collected using left/right ($\sigma$$^{-}$/$\sigma$$^{+}$) circular polarized x-rays beam with normal incidence at (e) 300 K under OOP magnetic field of 6 T and at (g) 150 K under OOP remanence (0 T). The corresponding XMCD signal is shown in (f) and in the bottom panel of figure (g).}
\label{FigS4}
\end{figure*} 

To get more idea about the average contribution of Fe$^{3+}$ and Fe$^{2+}$ cations at O$_{h1}$, O$_{h2}$ and T$_{d}$-sites, the Fe-L$_{2, 3}$ XMCD signal is modelled using CTM4XAS software~\cite{stavitski2010ctm4xas}. The initial parameters and fitting steps for the modelling are employed from the literature available for the pristine nickel ferrite~\cite{Lumetzberger2020, klewe2014physical}. The crystal field splitting 10 Dq = 1.6 eV and 10 Dq= -0.8 eV are used for Fe$^{3+}$ cations occupying the O$_{h}$ and T$_{d}$, respectively. Modelling results for the pristine, 3\% and 5\% Dy-doped sample are shown in Fig. S4b, Fig. S4c and Fig. S4d, respectively. From the fitting the contribution of Fe$^{3+}$, Fe$^{2+}$ at different crystallographic sites is obtained and shown in the inset of Fig. S4d, which reveals that contribution of Fe$^{2+}$ at O$_{h1}$-site is increased while contribution of Fe$^{3+}$ at O$_{h2}$ and T$_{d}$-sites decreases with increasing Dy concentration to 5\%. These results indicate a comparatively larger Dy$^{3+}$ ion moves Fe$^{3+}$ ions out of the O$_{h2}$ and T$_{d}$-sites or the majority of Dy$^{3+}$ occupying the O$_{h2}$-site (which is confirmed by Dy-L$_{3}$ EXAFS, discussed in context of Fig. 4 in the main text), leading to redistribution in the XMCD intensity.

\textbf{6.	X-ray Resonance Magnetic Reflectivity (XRMR)}

Reflectivity is measured in the specular mode before resonance energy and at the Fe/Ni-L$_{3}$ and Dy-M$_{5}$ edges in the longitudinal geometry with and without magnetic field for the 5\% Dy-doped NiZAF thin film. The data is recorded at fixed applied magnetic field (or without field) with changing the beam polarization. The signal is labelled with I$^{+}$ and I$^{-}$ recorded using right and left circular polarized x-rays beam, respectively. The considerable difference between I$^{+}$ and I$^{-}$ at the resonance indicates the hidden magnetic contrast~\cite{tonnerre2012depth} which can be seen, for example, in the reflectivity at 50 K for the Fe-L$_{3}$ edge under remanence state and at the Ni-L$_{3}$ edge reflectivity under 2 T shown in Fig. S5a and Fig. S5b, respectively. To extract the details of the magnetic information hidden in the magnetic reflectivity data, the magnetic asymmetry (R) is obtained using expression R= (I$^{-}$ - I$^{-}$)/(I$^{+}$ + I$^{-}$) , which is shown in Figs. S5c-5f. No considerable R was observed at Dy-M$_{5}$ edge (Fig. S5f), therefore we focus the asymmetry analysis on the Fe, Ni-L$_{3}$ edge. It is evident from Fig. S5c for the remanence conditions at 300 K that R is nearly zero at all momentum transfer q-values, but has a finite negative value over the entire q-range including some oscillations at 50 K. Furthermore, R value increases considerably under an external IP field of 2 T at 50 K compared to the remanence state due enhancement in the magnetic ordering in presence of external field (see Fig. S5d). Since R has a finite non-zero value at maximum q-value [q$_{max}$ = 5 nm$^{-1}$; see Fig. S5c-5d] at 50 K, it indicates the presence of OOP magnetization component in addition to the IP component in the sample at 50 K in contrast to the 300 K. This can be understood qualitatively in view of the m$_{z}$ sin($\theta$) term in the atomic scattering factor~\cite{tonnerre2012depth, violbarbosa2012inhomogeneous, meyerheim2009new}. We note that R is nearly zero in the remanence state at 300 K due to smaller magnitude of magnetization compared to low temperature of 50 K. Moreover, at 300 K, R is zero at q$_{max}$ = 5 nm$^{-1}$ indicating no OOP magnetic component which is expected since such pristine NiZAF has been reported to host IP magnetic anisotropy~\cite{Lumetzberger2020, Emori2017}.

The magnetic reflectivity and asymmetry are modelled using Dyna software~\cite{elzo2012x}. The obtained fitting of reflectivity and asymmetry are given in Fig. 6 and Fig. 7, respectively. To fit the reflectivity and asymmetry data, we used XAS/XMCD data collected at 150 K for the Ni, Fe-L$_{2, 3}$ and Dy-M$_{4, 5}$ edges to obtain the imaginary part of scattering factors/optical constant~\cite{tonnerre2012depth}. Chantler table~\cite{chantler2000detailed} is employed to obtain the scattering factors for elements Zn, Ni and O for which we did not measured XAS/XMCD. Subsequently, real part of scattering factors is calculated using Kramers-Kronig relationship~\cite{tonnerre2012depth}.

\begin{figure*}[t]
\center
\includegraphics[width=1\linewidth]{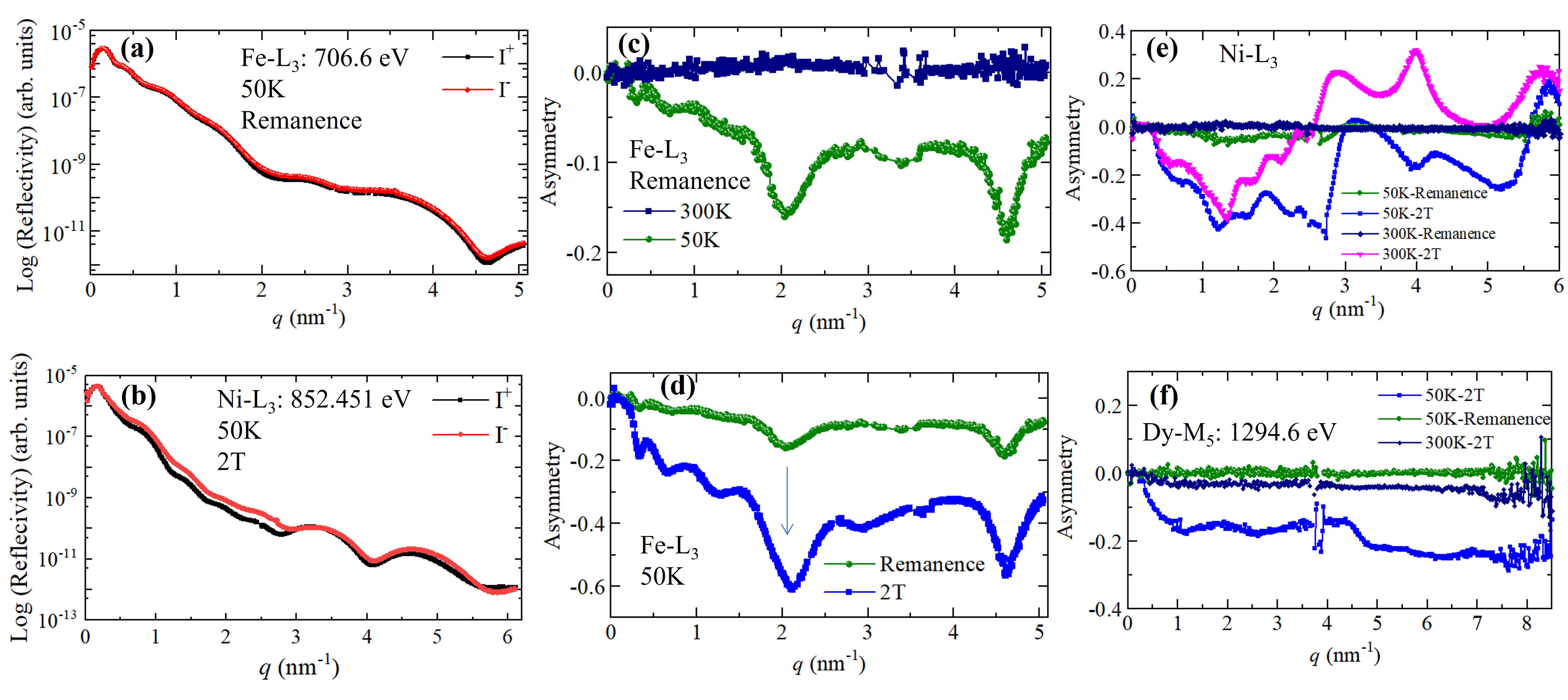}
\caption{\textbf{XRMR and magnetic asymmetry for 5\% Dy-doped NiZAF film.} Reflectivity verses momentum transfer (q) collected at 50 K for (a) Fe-L$_{3}$ in remanence and (b) Ni-L$_{3}$ at 2 T. I$^{+}$ and I$^{-}$ corresponds to the reflectivity measured using right and left circular polarized x-rays, respectively. (c) Magnetic asymmetry R=(I$^{-}$ - I$^{-}$)/(I$^{+}$ + I$^{-}$)for Fe-L$_{3}$ in the remanence state at 300 K and 50 K. (d) Comparison of R in the remanence with R under 2 T for Fe-L$_{3}$ at 50 K, where arrow is a guide to eye to see significant enhancement in the magnitude of R under 2 T compared to remanence state. R measured under 2 T and in remanence at 300 K and 50 K for (e) Ni-L$_{3}$ and (f) Dy-M$_{5}$.}
\label{FigS5}
\end{figure*}

\begin{figure*}[t]
\center
\includegraphics[width=1\linewidth]{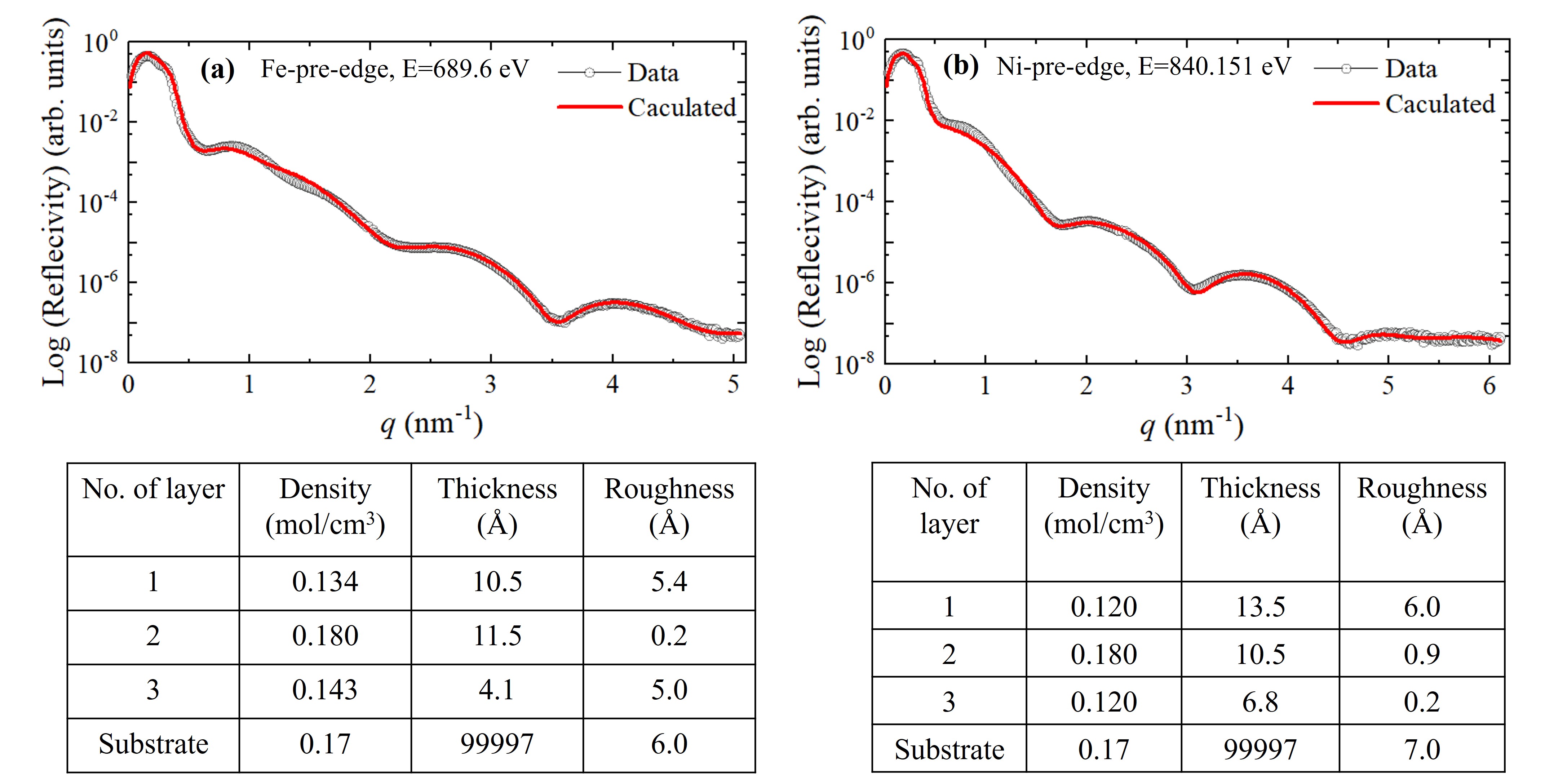}
\caption{\textbf{Modelling of the magnetic reflectivity for the 5\% Dy-doped NiZAF film}. Calculated magnetic reflectivity (red curve) using the data recorded with right circular polarized x-rays at 50 K for (a) energy E = 689.6 eV before Fe-L$_{3}$ resonance and (b) energy E = 840.151 eV before Ni-L$_{3}$ resonance. Geometrical parameters obtained from the modelling are shown in table given below of respective figures.}
\label{FigS6}
\end{figure*}

\begin{figure*}[t]
\center
\includegraphics[width=1\linewidth]{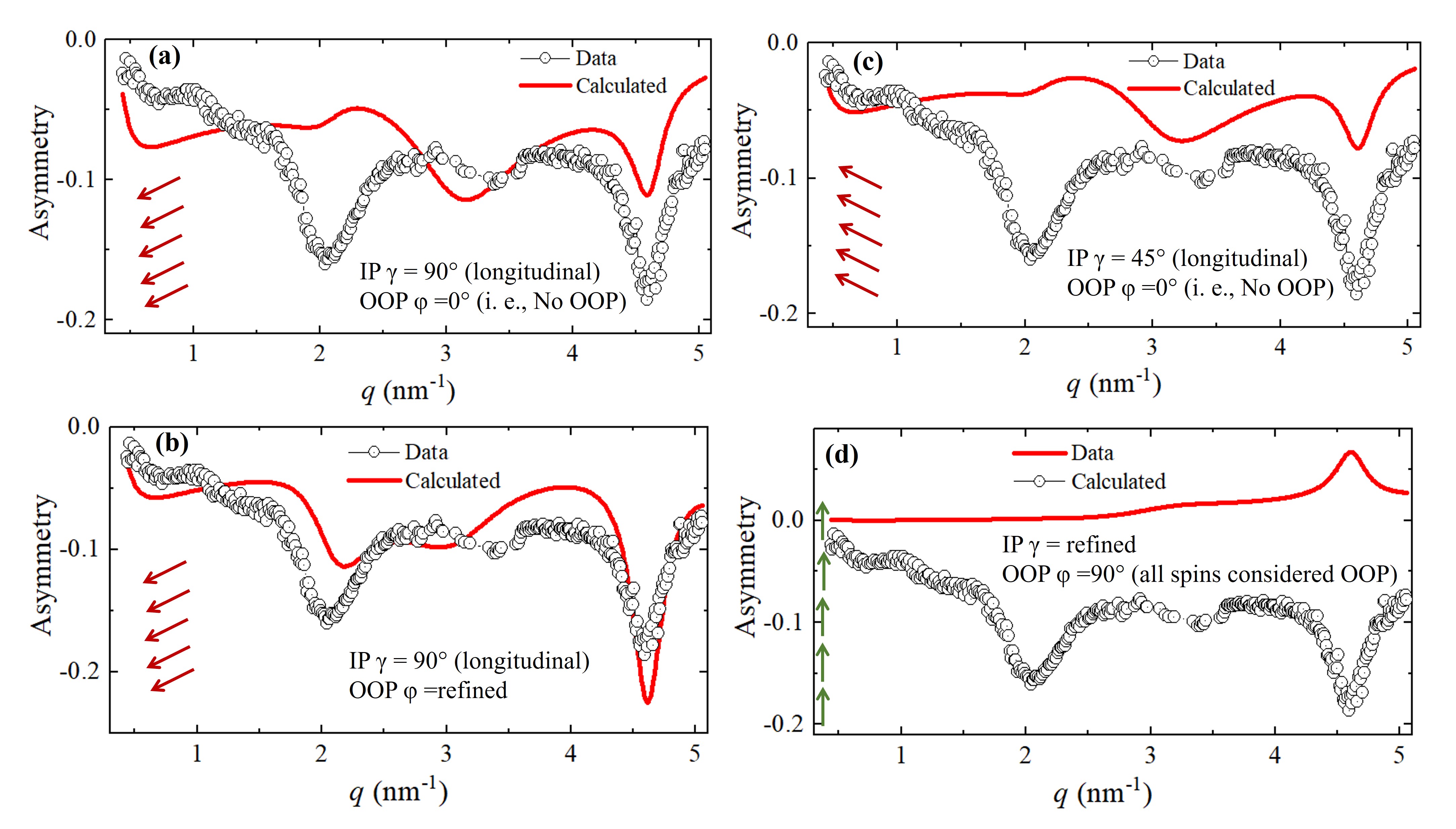}
\caption{\textbf{Magnetic asymmetry analysis for the 5\% Dy-doped NiZAF film}. Calculated magnetic asymmetry (red curve) using data collected at 50 K in the remanence state at the Fe-L$_{3}$ edge for different magnetic models of (a) $\gamma$ = 90°, $\phi$ = 0°, (b) $\gamma$ = 90°, $\phi$ is refined, (c) $\gamma$ = 45°, $\phi$ = 0° and (d) $\gamma$ is refined, $\phi$ = 90°. The schematic of the corresponding spin structure for different layers (top-to bottom: total 5 layers) are indicated by arrows in the inset of each figure. None of these models are able to capture the experimental asymmetry data.}
\label{FigS7}
\end{figure*}

\begin{figure*}[t]
\center
\includegraphics[width=1\linewidth]{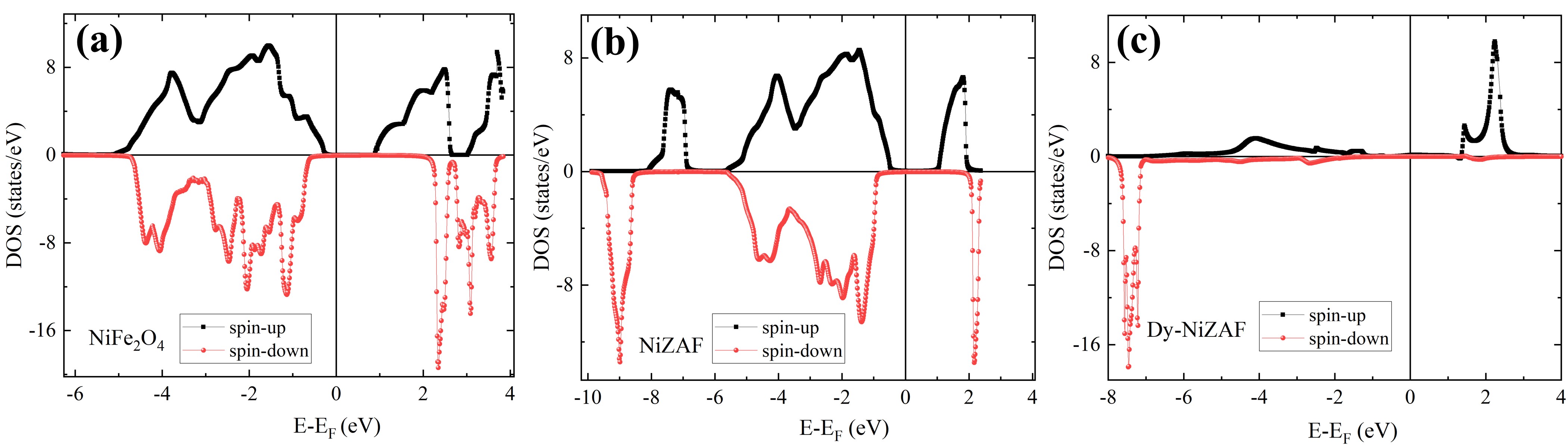}
\caption{Density of States (DOS) for (a) NiFe$_{2}$O$_{4}$, (b) NiZAF and (c) Dy-doped NiZAF. }
\label{FigS8}
\end{figure*}

In the first step, the reflectivity data is modelled for energy far before the resonance to extract the geometrical structure for 5\% Dy-doped NiZAF. For instance, the fitting using the reflectivity data measured with right polarized x-rays at 50 K for energy E = 689.6 eV far before the Fe-L$_{3}$ resonance (E = 706.6 eV) and energy E = 840.151 eV far before the Ni-L$_{3}$ resonance (E = 852.451 eV) are shown in Fig. S6a and Fig. S6b, respectively. For each layer, density of the Dy-doped NiZAF film is varied within ±10-15\% from the experimental value. Form the modelling of reflectivity, we find that a minimum 3-layeres of Dy-doped NiZAF along with substrate layer provide a reasonable fit to the experimental reflectivity data (see Fig. S6). We note that one contamination layer (refined by using a carbon layer) with thickness around 2 nm was needed to be included in the modelling on the surface of the samples which is usual in practice due to surface exposure of the films in different environment (beam exposure, air exposure, temperature training 2-400 K, etc.). The geometrical parameters (density, thickness, roughness) obtained from the modelling are shown in below of respective Figs. S6a and S6b. The total thickness of Dy-doped NiZAF estimated around 2.6(4) nm for Fe and 3.1(4) nm for Ni from the corresponding reflectivity analysis. This subtle difference in the total thickness for Fe and Ni is within the statistical/systematic errors. However, a possible reason for this difference in the thickness would be, since Ni and Fe have different chemical environments in the present sample (Ni occupies octahedral sites while Fe occupies both tetrahedral and octahedral sites) some difference in total thickness between Fe and Ni might be expected in the x-ray reflectrometry due to chemical contrast through x-ray form factor~\cite{tonnerre2012depth}. Nevertheless, this is not important and rather we focus here on the magnetic profile.

The geometrical parameters are found to be nearly similar using the reflectivity data collected at the energy far before the resonance (before Fe, Ni-L$_{3}$ edge i.e., off-resonance) and at the resonance energy (Fe, Ni-L$_{3}$). However, a moderate difference in the overall behavior of reflectivity at the off-resonance and at the resonance is expected (see Figs. S5a-S5b and S6a-S6b) due to change in atomic scattering factors at the resonance energy~\cite{krieft2020advanced}. After obtaining the geometrical parameters form the average reflectivity I$_{avg}$= (I$^{+}$+I$^{-}$)/2 analysis at resonance, magnetic asymmetry (R) is modelled in the next step as shown in Fig. S7 and Fig. 3 of the main text. To fit the asymmetry data measured at the Fe-L$_{3}$ edge at 50 K under a 2 T magnetic field applied at the intersection of the scattering and sample planes (longitudinal direction), we used scattering factors derived from XAS/XMCD experiments collected at 150 K under 6 T. The R is modelled using the different magnetic models keeping the geometrical parameters unchanged to those values obtained from the reflectivity refinements. Different magnetic models were tested to fit the experimental asymmetry data by varying the IP azimuthal angle ($\gamma$) and OOP polar angel ($\phi$) angles are varied [see Fig. 3f of the main text for visualizing the angles $\gamma$ and $\phi$]. The overall shapes of the R depend on $\gamma$ and $\phi$ while its magnetic usually depend on the fraction of the magnetic moment value at the particular energy~\cite{tonnerre2012depth}. To model the experimental R for the Fe-L$_{3}$ in the remanence sate, several different collinear and noncollinear magnetic structure were tried which failed to provide a satisfactory fit the data, for instance, some of those spin profile and corresponding fits are shown in Fig. S7a-S7d. The best magnetic profile which could provide a reasonable fit with corresponding magnetic angles ($\gamma$ and $\phi$) are shown in Fig. 3c and Table I of the main text. Similar protocols were used to model the R data measured under 2 T for Fe-L$_{3}$ and Ni-L$_{3}$ resonance. The corresponding best fits with related discussion are given in Fig. 3d,e of the main text.

\par

\medskip

\bibliographystyle{apsrev}
\bibliography{./Dy-NIZAF}
\small